\documentclass[nonatbib,preprint,12pt]{elsarticle}

\usepackage[
backend=biber,
style=numeric,
sorting=none,
maxcitenames=50,
maxnames=50
]{biblatex}

\usepackage{amssymb}
\usepackage{amsmath}
\usepackage{bm}
\usepackage{bbm}
\usepackage{amsthm}
\usepackage{dsfont}
\usepackage{hyperref}
\usepackage{xcolor}
\usepackage{multirow}

\DeclareUnicodeCharacter{02B9}{'}

\begin{document}

\sloppy

\begin{frontmatter}
	
	\title{ Multilevel radial basis function surrogates for noise-robust DSMC-CFD coupling }
	
	\author[label1]{Arshad Kamal\footnote{ arshad.kamal10@alumni.imperial.ac.uk. Author for correspondence. } } \author[label1]{Arun Kumar Chinnappan\footnote{ a.chinnappan@soton.ac.uk} } \author[label2]{James R. Kermode\footnote{ j.r.kermode@warwick.ac.uk}} \author[label1]{Duncan A. Lockerby\footnote{ d.lockerby@warwick.ac.uk}} 
	
	\affiliation[label1]{organization={School of Engineering, University of Warwick},
		addressline={Library Road}, 
		city={Coventry},
		postcode={CV4 7AL}, 
		country={United Kingdom}}
	
	\affiliation[label2]{organization={Warwick Centre for Predictive Modelling, School of Engineering, University of Warwick},
		addressline={Library Road}, 
		city={Coventry},
		postcode={CV4 7AL}, 
		country={United Kingdom}}
	
	\begin{abstract}
		Hybrid methods for simulating rarefied gas flows reduce computational cost by coupling a particle-based model, typically the direct simulation Monte Carlo (DSMC) method, to a continuum-based solver, i.e. a computational fluid dynamics (CFD) code. However, widespread adoption of these methods is hindered by numerical instabilities caused by statistical noise and difficulties in applying them to complex geometries. To be effective, a hybrid framework must be robust to noise, reliable in not introducing errors to the flow physics, automated, and flexible enough for complex spatial domains. Previous iterations of the micro-macro-surrogate-sparse (MMS-Sparse) method successfully addressed the first three requirements using Bayesian surrogate models to provide smooth, constitutive corrections to the CFD. However, they relied on global basis functions, limiting their applicability to relatively simple geometries. In this work, we address the fourth requirement - flexibility - by introducing a set of multilevel radial basis functions (RBFs) to represent the smooth corrections within the MMS-Sparse framework. Unlike global polynomials, multilevel RBFs can resolve broad and fine flow details locally, potentially allowing the method to be applied to more complex geometrical systems than previously considered. We couple this approach with a finite-volume CFD solver (OpenFOAM) and validate it using the rarefied lid-driven cavity flow problem. This serves as a rigorous test case for spatially two-dimensional coupling. Our results demonstrate that this enhanced MMS-Sparse method produces estimates in good agreement with benchmarks while retaining the noise-robust and automated benefits of the Bayesian approach.
	\end{abstract}
	
	\begin{keyword}
		low-speed rarefied gas flows \sep DSMC-CFD coupling \sep micro-macro-surrogate-sparse  \sep sparse Bayesian learning \sep multilevel radial basis functions \sep OpenFOAM
	\end{keyword}
	
\end{frontmatter}

\section{Introduction}

\label{sec:introduction}

The study of rarefied gas flows is important in numerous industrial applications across various spatial and time scales, such as in the modelling of atmospheric re-entry in aerospace, or in the design of micro-electromechanical systems (MEMS) in microfluidics. A gas can be referred to as being `rarefied' when the mean free path, $\lambda$, of the molecules is large in comparison to the characteristic length, $L$, of the system \cite{landau2013general}. As such, one can associate a non-dimensional quantity, the Knudsen number $Kn$, to a rarefied gas flow:

\begin{equation}
	Kn = \lambda / L. \label{eq:KnudsenNumber}
\end{equation}

Here, the mean free path $\lambda$ denotes, on average, the distance travelled by a gas molecule between two successive collisions, while $L$ is a quantity that signifies the scale of the system. For example, on a large scale, of order meters, the flow past the spacecraft in atmospheric re-entry at higher altitudes has a comparatively large mean free path compared to the characteristic length of the system (i.e. $\lambda \gg L$). On much smaller scale, in the case of MEMS devices, the characteristic length scale is of order micrometres to millimetres and is comparatively small compared to the mean free path (i.e. $L \ll \lambda$). Correspondingly, in both examples, on opposite scales in length, the Knudsen number $Kn$ can be significant (i.e. $Kn \gg  1$) and the gas can be rarefied. \\

As one increases $Kn$ from $0$, we move across four gas regimes \cite{schaaf1961flow}, where flow becomes more `particle-like' and rarefied. In the first regime, with $Kn < 0.001$, the flow can be modelled as a `continuum' fluid, where the Navier-Stokes (N-S) equations can be applied with no-slip boundary conditions. Increasing to $0.001 < Kn < 0.1$, the N-S equations, modified with appropriate slip boundary conditions, can be used to accurately capture the macroscopic properties. However, as we increase $Kn$ even further to $0.1 < Kn < 10$, the accuracy of the N-S equations, even with slip boundary conditions, becomes limited. In this `transitional' regime, and particularly so in the `free molecular' regime $Kn > 10$, the Boltzmann equation, or some approximation, must be adopted. Although the Boltzmann equation is applicable across all four gas flow regimes, it is extremely difficult to solve analytically or numerically, except in the free molecular regime, due to its complex collision integral term. Therefore, the direct simulation Monte Carlo (DSMC) method \cite{bird1994molecular}, a kinetic-theory-based, stochastic technique that directly simulates the Boltzmann equation, is typically employed in the transitional and free molecular regimes. DSMC simulates representative particles, where each simulated molecule represents a group of real gas molecules. Molecular movement is treated deterministically, while collisions are modelled probabilistically using an acceptance-rejection algorithm. Macroscopic properties are extracted from microscopic data through time-averaging for steady-state solutions and ensemble-averaging for unsteady cases. Since DSMC inherently provides solutions consistent with the Boltzmann equation, it is, in principle, valid across all regimes. However, for accuracy, the DSMC algorithm requires that the computational cell size be smaller than the local mean free path, with at least 15–20 simulated particles per cell, and that the time step be smaller than the mean collision time. In the slip and continuum regimes, this leads to a large number of very small cells, a reduced time step, and a significantly increased number of simulated particles; thereby greatly increasing the computational cost. Additionally, the statistical nature of the method necessitates a significant number of samples to achieve results with low statistical noise, particularly at low fluid velocities. For instance, the number of samples needed for low-noise results scales approximately with the inverse square of the local flow Mach number \cite{hadjiconstantinou2003statistical}. These factors make this method computationally challenging in the transitional and late slip regimes and practically prohibitive in the continuum regime for low-speed flows. \\

The high computational burden of the DSMC method is a strong motivation for the use of hybrid domain decomposition methods, where the domain of the physical problem is decomposed into regions of near-equilibrium and non-equilibrium. For example, in the case of microchannel flows in the slip regime \cite{karniadakis2006microflows}, regions close to the boundaries can be in non-equilibrium due to the presence of `Knudsen layers', while, in the bulk, the flow is in near-equilibrium. Decomposing the domain, particle-based solutions to the Boltzmann equation can be obtained in the Knudsen layers using, for example, the DSMC method \cite{bird1994molecular}, and continuum methods to solve the N-S equations can be employed in the remaining near-equilibrium regions. Such hybrid domain decomposition methods are widely documented in the literature \cite{wadsworth1992two} \cite{o1995molecular} \cite{garcia1999adaptive} \cite{schwartzentruber2006hybrid} \cite{pantazis2014hybrid} \cite{docherty2016coupling} \cite{john2018simulation} \cite{zhang2019particle} \cite{vasileiadis2024hybriddcfoam} \cite{kumar2024cartesian} \cite{nompelis2025verification}. \\

While such hybrid coupling methods present an efficient solution for simulating rarefied gas flows at high speeds, coupling at low speeds presents unique challenges, preventing their general application to arbitrary flow conditions. The first challenge relates to the transfer of noisy data between the particle-based solver and the continuum-based CFD model, which can lead to numerical instabilities in the coupling algorithm \cite{docherty2016coupling} \cite{pantazis2014hybrid}. While the DSMC noise can be reduced by increasing the number of particles per cell and/or the sample window, the increased computational cost associated with this can outweigh the benefits of adopting a hybrid approach in the first place. While one may instead use simple filtering/averaging strategies for denoising, such approaches do not require that information transferred to the CFD satisfy physical conservation equations, nor do they guarantee that the filter parameters are not manually tuned or do not spuriously affect the overall solution. Another challenge to coupling at low speeds specific to steady-state simulations is running the DSMC solver alongside the CFD until a converged solution is achieved across their interface. This coupling can be computationally expensive, particularly in 2D/3D simulations, due to the cost of DSMC. In addition to these two challenges relating to noise, there is also the challenge of applying DSMC-CFD to complex geometries and flow configurations. To overcome all the above challenges, we seek a DSMC-CFD coupling that satisfies the following four requirements: (i) robustness in reducing DSMC noise and computing smooth corrections, (ii) reliability in satisfying constitutive conservation equations, (iii) automation in obtaining CFD estimates, (iv) flexibility in application to two- and three-dimensional domains. \\

To meet these requirements, we employ an existing hybrid DSMC-CFD coupling framework based on the micro-macro-surrogate (MMS) method, originally proposed by Tatsios et al. \cite{tatsios2025dsmc} for steady-state Poiseuille flows and extended for sparsity (MMS-Sparse) by Chinnappan et al. \cite{chinnappan2025bayesian} for unsteady Poiseuille flows. In this hybrid, surrogate models for macroscopic quantities are placed in between the micro (a particle-based solver) and macro (a continuum-based solver) models. These form surrogates to the particle-based solutions and are used to produce noise-free, constitutive corrections for continuum-based solvers to form robust macro-model estimates. The surrogate model in this hybrid reduces statistical noise and eliminates the need for full DSMC simulations in non-equilibrium regions. Instead, stress corrections and slip boundary conditions (estimated from a limited number of DSMC simulations) are added to the CFD solver. Importantly, the surrogate models are not introduced as a DSMC noise-reduction technique, such as Proper Orthogonal Decomposition (POD) \cite{kumar2018denoising}, which does not explicitly enforce conservation. Instead, the surrogates are used to estimate constitutive corrections that enter directly into the governing equations, thereby ensuring that the resulting flow field remains consistent with the fundamental conservation principles. The entire domain is eventually solved using CFD: standard CFD in equilibrium regions, and CFD with corrections in non-equilibrium regions once the required information has been extracted from the limited DSMC data. Although such a surrogate modelling approach has been applied to multi-scale fluidic systems \cite{stephenson2018accelerating}, we employ this in the context of multi-scale, low-speed rarefied gas flows. In addition, the surrogates are learned using sparse Bayesian learning. While other methods based on Neural Networks may be employed for the surrogates \cite{zhang2025shape} \cite{roohi2025data} \cite{roohi2026accelerating}, the advantages of a Bayesian approach is that it is less prone to over-fitting on noisy datasets when compared to frequentist approaches based on maximum likelihood estimates and provides an automatic determination of model complexity \cite{bishop2006pattern}. \\

We use the existing MMS-Sparse methodology from Chinnappan et al. \cite{chinnappan2025bayesian} to meet the aforementioned requirements for a DSMC-CFD coupling. While this existing methodology satisfies three of the requirements for a DSMC-CFD coupling, namely robustness, reliability and automation, the novelty of this work is from the use of multilevel radial basis functions (RBFs) for flexibility. This flexibility requirement represents a significant methodological step because spatially two-dimensional problems can provide much more flow complexity than the spatially one-dimensional cases considered previously. Our approach with multilevel RBFs is an automated approach that gives the ability to flexibly capture both broad and fine details of flow configurations as and when necessary. While sophisticated, our multilevel RBFs approach can compute credible intervals automatically for macroscopic quantities of interest through sparse Bayesian learning. In addition to the use of multilevel RBFs, our study further differs from Chinnappan et al. \cite{chinnappan2025bayesian} in that corrections for the macro-model are computed in two-dimensions in space and an open-source, finite-volume method (i.e. OpenFOAM) is used to solve the governing equations of the macro-model.  For our spatially two-dimensional demonstration of MMS-Sparse, we use the steady-state lid-driven cavity (LDC) flow problem as the test case due to the extensive prior work on this classical problem \cite{naris2005driven} \cite{john2010investigation} \cite{wu2016sound} \cite{wang2018oscillatory} \cite{wang2018nonlinear} \cite{wang2019heat} \cite{he2024numerical}. This is the natural and logical first test case for a two-dimensional problem after previous work on spatially one-dimensional flows and is significantly more complex. Unlike many of the previous studies on the LDC, we consider mainly low-speed flows and use the problem merely as a test case to assess the performance of our hybrid DSMC-CFD approach for a spatially two-dimensional problem. \\

In this work, we focus solely on the DSMC method as our `micro model'. However, we note that a host of DSMC such as low-variance DSMC \cite{baker2005variance} \cite{homolle2007low}, variance-reduced DSMC \cite{al2010low} \cite{sadr2023variance} and Fokker-Planck variants \cite{gorji2015variance} \cite{sadr2023variance2} have been developed for reducing statistical noise specifically in the low-speed limit. Alternative particle-based methods to the DSMC, such as that based on Bhatnagar-Gross-Krook (BGK) \cite{naris2005driven}, may also be used in the low-speed limit. Recent work has explored such alternative methods \cite{fei2023time} \cite{cui2025multiscale} \cite{luo2025boosting} \cite{yang2025denoising}, which includes a denoising multiscale particle (DMP) method \cite{yang2025denoising} that simplifies the binary collision process through a BGK relaxation and incorporates a strategy similar to the information preservation (IP) \cite{fan2001statistical} method originally introduced in the DSMC method. The rational for adopting the standard DSMC method in our hybrid method for the low-speed case is so that a generally applicable hybrid solution might be obtained across all flow regimes. Demonstration of the applicability of our hybrid DSMC-CFD approach to compressible flows will be published in future work.  \\ 

We organise this paper as follows. In Section \ref{sec:mms-sparse}, we detail the MMS-Sparse hybrid method for a spatially two-dimensional problem, in particular the choice of surrogate models for the flow velocity and stress tensor. Section \ref{sec:sparse-bayesian-learning} discusses the robust sparse Bayesian learning approach to learn surrogate models and incorporate smooth corrections. Section \ref{sec:multilevel-radial-basis-functions} details the flexible, multilevel RBFs used as the basis functions for the sparse Bayesian learning. We then assess the MMS-Sparse method for the spatially two-dimensional LDC test case. In Section \ref{sec:proof-of-concept}, we first present a proof-of-concept of the method by using, as input, particle-based solutions for the entire domain. Then, in Section \ref{sec:pseudo-hybrid-approach}, we test the method in a more realistic framework, where only particle-based solutions in a region $2\lambda$ from the walls are used as input to obtain flow estimates for the entire domain, as one may typically encounter in a hybrid coupling algorithm. Finally, Section \ref{sec:discussion} forms a discussion and sets proposals for future work. 

\section{MMS-Sparse hybrid for a two-dimensional cavity}

\label{sec:mms-sparse}

\begin{figure}
	\centering
	\includegraphics[scale=0.25]{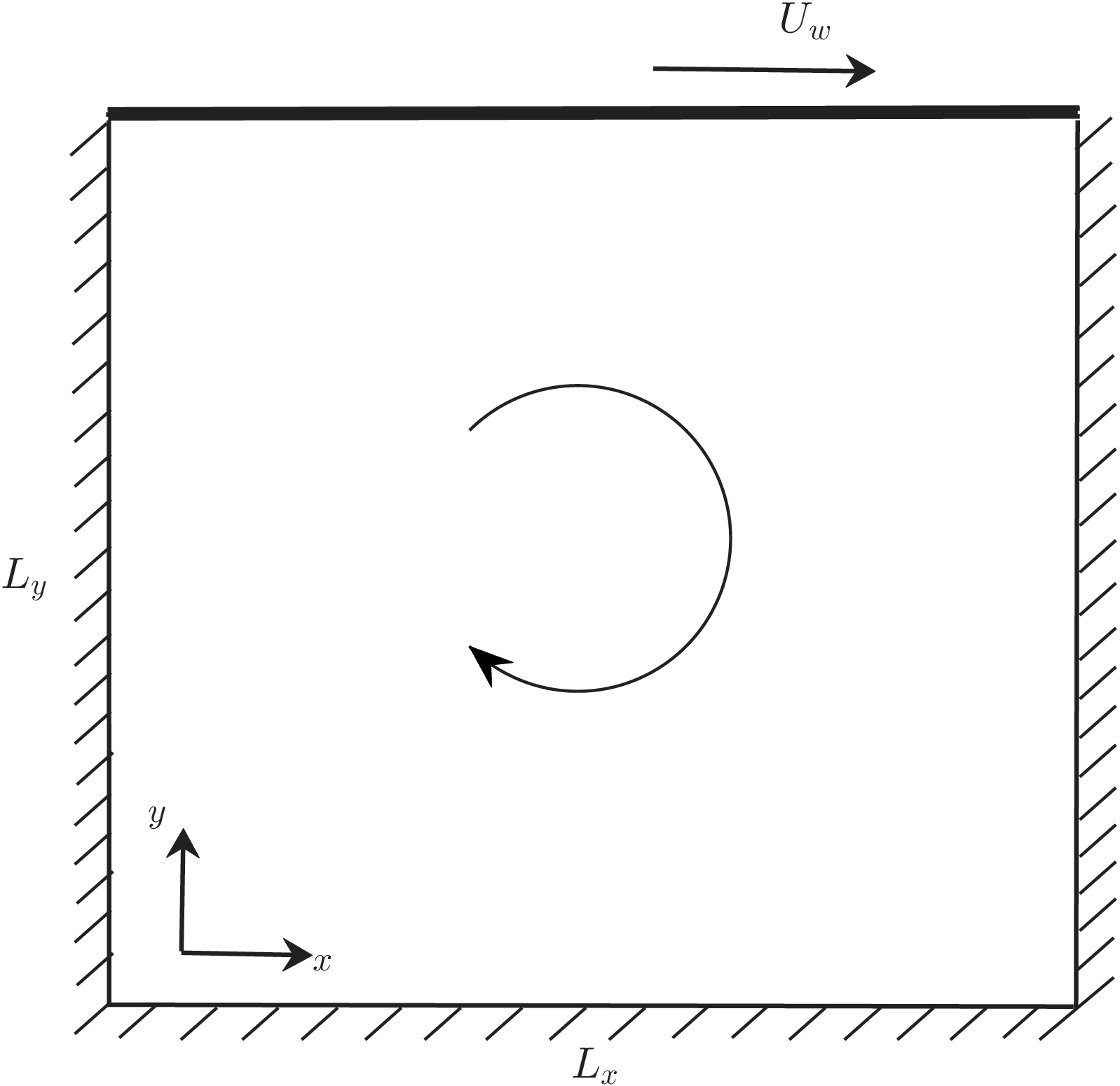}
	\caption{ The geometry of the 2D lid-driven cavity problem, where $L_{x}$ and $L_{y}$ are the lengths of the walls and $U_{w}$ is the velocity which the upper wall is initialised within the DSMC simulation. }
	\label{fig:LDC_Geometry}
\end{figure}

The micro-macro-surrogate (MMS) method was originally introduced by Tatsios et al. \cite{tatsios2025dsmc}, where it was demonstrated for steady-state Poiseuille flows. Along with a proof-of-concept of the method using particle-based data, they presented promising results in a hybrid framework, where only DSMC data close to the walls was used as input to obtain accurate flow estimates for the entire domain. They also formed comparisons with traditional hybrid methods. This initial work, however, relied upon ad hoc exponential basis functions to learn the smooth corrections for the CFD solver. \\

To move away from such an ad hoc formulation, the MMS methodology was extended by Chinnappan et al. \cite{chinnappan2025bayesian} using an automated procedure, based on the fast `Sequential Sparse Bayesian Learning' (SSBL) algorithm \cite{tipping2003fast} and complete Chebyshev-Fourier basis functions suitable for bounded flows. This extended formulation, termed MMS-Sparse, was demonstrated for unsteady Poiseuille flows, with the in-principle computational savings presented for a broad range of cases. \\

In the present work, we extend the MMS framework further by introducing multilevel, squared-exponential RBFs to learn the smooth corrections. This approach eliminates the need for ad-hoc or problem-specific basis selection and enhances the flexibility and scalability of the MMS framework, making it potentially applicable to complex flow configurations. As a test case, we consider a more complex configuration: the two-dimensional lid-driven cavity flow (see Figure \ref{fig:LDC_Geometry}). This is the logical next step for two-dimensional problem, after previous consideration of spatially one-dimensional flows. This canonical flow features significant flow recirculation, multidirectional (spatial) gradients and complex vortex structures, representing a substantial departure from the spatially one-dimensional channel flows considered in previous studies. Moreover, the lid-driven cavity flow is a classical benchmark problem for validating numerical methods due to its sensitivity to the coupling algorithm and boundary conditions. \\

\begin{figure}
	\centering
	\includegraphics[scale=0.28]{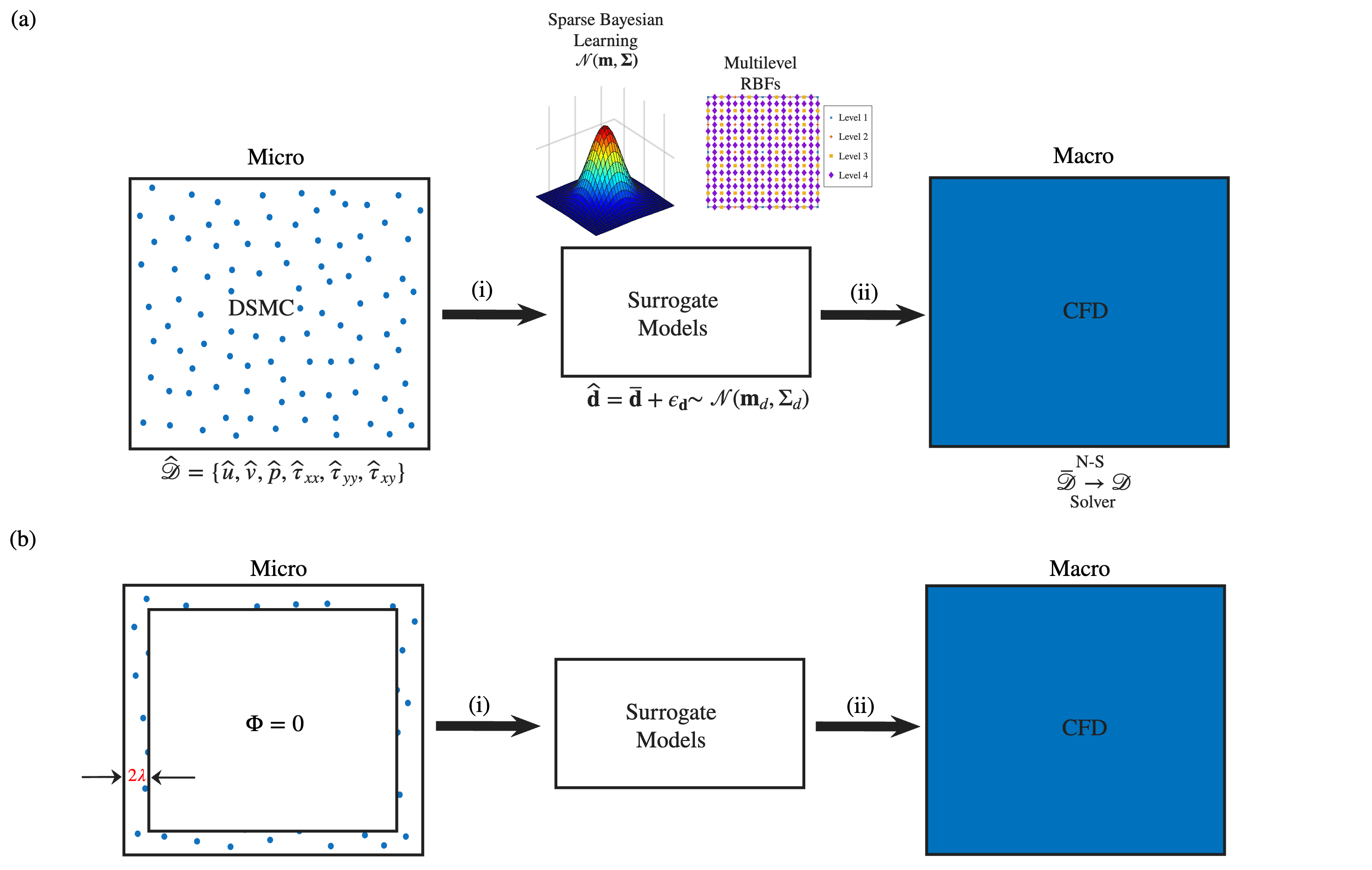}
	\caption{ (a) Proof-of-concept of MMS-Sparse. (i) Noisy DSMC data produced from micro-model simulation over the entire domain is used to obtain surrogate models via Bayesian inference and multilevel RBFs. (ii) Smooth corrections are calculated from the surrogate models and embedded within a macro-model CFD to form estimates on the flow. (b) Pseudo-hybrid approach with MMS-Sparse. (i) Only noisy DSMC data within $2\lambda$ of each wall is used by the surrogate models, with zero corrections interior ($\Phi = 0$). (ii) Corrections are computed from the surrogate models, assuming zero corrections interior, and used inside macro-model CFD solver to obtain estimates.  }
	\label{fig:MMS-Sparse_OneWayCoupling}
\end{figure}

Figure \ref{fig:MMS-Sparse_OneWayCoupling} shows, in relation to the LDC flow problem in a one-way coupled framework, the (a) proof-of-concept and (b) pseudo-hybrid demonstrator of the MMS-Sparse method. The proof-of-concept is for verification to ensure that, when DSMC data for the entire domain is provided as input, the hybrid approach works as intended compared to the benchmark. The pseudo-hybrid approach is used to demonstrate the potential computational savings that can be achieved when only data close to the walls is collected and provided as input in MMS-Sparse. \\ 

As one can see from Figure \ref{fig:MMS-Sparse_OneWayCoupling}, there are three components to MMS-Sparse hybrid: the micro model (the DSMC method), the macro model (OpenFOAM; an open-source finite-volume CFD code), and the surrogate models inferred using a sparse Bayesian method and multilevel radial basis functions. In the subsequent subsections, we detail each of these components. 

\subsection{Micro model}

For the micro-model, we use the direct simulation Monte Carlo (DSMC) method \cite{bird1994molecular}. Although other kinetic models, such as those based on the Bhatnagar-Gross-Krook (DSBGK) model, S-model or ES-model, may instead be employed \cite{naris2005driven} \cite{li2011direct} \cite{ho2019comparative} \cite{fei2020unified} \cite{fei2025navier}, the DSMC method has the advantage of being accurate for all Knudsen numbers, particularly at high-speeds as it gives solution to the full Boltzmann equation without any approximation for the collisional term. In the DSMC method, ballistic motion is used to update the particle positions within the domain. Particles that encounter solid surfaces during this update have their motion modified according to prescribed gas-surface interaction models. In this work, diffuse reflection with full thermal accommodation is employed. Particle–particle collisions are considered separately using the variable hard sphere (VHS) model \cite{bird1981monte}. Based on both particle-surface and particle-particle collisions, microscopic quantities, such as particle velocity and shear stress, can be computed. Since the DSMC is a stochastic method, there can be significant noise when computing these quantities, although the level of noise can be reduced by simulating over a long duration and/or by averaging over many independent runs. \\

We utilise an open-source solver, the Stochastic PArallel Rarefied-gas Time-accurate Analyzer (SPARTA) \cite{plimpton2019direct}\cite{SPARTA-DSMC}, to perform the DSMC. Here, Message Passing Interface (MPI) is used to parallelise the DSMC simulations via domain decomposition. Particle-surface and particle-particle collisions are calculated for each cell, from which the particle velocity, scalar pressure and stress is calculated per cell. Although not exploited in the present work, the domain decomposition nature of SPARTA allows for an ease of communication with continuum-based domain decomposition methods also using MPI, such as OpenFOAM \cite{weller1998tensorial}.  

\subsection{Macro-model}

\label{ssec:macro-model}

To ensure that the estimates from our hybrid approach are consistent with conservation principles, we supply smooth corrections to the N-S equations and their associated boundary conditions. To solve the N-S equations for the macro-model, we use the icoFoam solver \cite{greenshields2021} in OpenFOAM \cite{weller1998tensorial}. This is based on the Pressure-Implicit with Splitting of Operators (PISO) algorithm \cite{ferziger2002computational} with finite-volume discretisation (see \ref{sec:piso-algorithm} for more details on the PISO algorithm). The icoFoam solver is used to solve the following incompressible N-S equations, with the continuity equation,
	\begin{eqnarray}
		\frac{\partial {\bf u}}{\partial t} + ({\bf u} \cdot \nabla) {\bf u} - \nabla \cdot (\nu \nabla {\bf u}) + \frac{1}{\rho} \nabla \cdot \Phi = -\frac{1}{\rho} \nabla p, \label{eq:NavierStokesMomentum}\\
		\nabla \cdot {\bf u} = 0, \label{eq:NavierStokesContinuity}
	\end{eqnarray}
on a rectangular domain $\Omega = [-L_{x}/2, L_{x}/2] \times [-L_{y}/2, L_{y}/2]$. Here, ${\bf u} = (u, v)$ is the flow velocity, $\nu = \mu / \rho$ is the kinematic viscosity, $\mu$ is the dynamic viscosity, $\rho$ is the mass density, $p$ is the scalar pressure, and the rank-2 stress correction tensor $\Phi$ is given by 
\begin{equation}
	\Phi = \begin{bmatrix}
		\phi_{xx} & \phi_{xy} & 0 \\
		\phi_{yx} & \phi_{yy} & 0 \\
		0 & 0 & 0 \\
	\end{bmatrix},
\end{equation} 
whose components we shall define later in Section \ref{sec:sparse-bayesian-learning}. On the boundaries of this domain, which we denote by $\partial \Omega$, we impose Dirichlet boundary conditions: ${\bf u}\vert_{\partial \Omega} = \bar{\bf u}$ and $p\vert_{\partial \Omega} = \bar{p}$, where $\bar{\bf u}$ and $\bar{p}$ are smooth boundary corrections to be found from surrogate modelling of the microscopic model. 

\subsubsection{Assignment of boundary and stress corrections}

Since we are using a finite-volume method, given a cell within the computational domain, one needs to specify face values for the boundary and stress corrections. We assume that the cell-centred value, obtained from DSMC, can be used uniformly across the face. \\

For example, for the x-component of the flow velocity, $u$, on a boundary face $\partial \Omega_{i}$, a value can be assigned based on the cell-centres:
\begin{equation}
	u_{i} = \frac{1}{S_{i}} \int_{\partial \Omega_{i}} u({\bf x}) d{\bf x} \overset{!}{=}  \frac{1}{S_{i}} \int_{\partial \Omega_{i}} u^{c_{i}} d{\bf x} = u^{c_{i}},
\end{equation}
where $S_{i}$ is the surface area of the boundary face $\partial \Omega_{i}$ under consideration and $u^{c_{i}}$ is the cell-centred value of $u$ on $\partial \Omega_{i}$ obtained from the surrogate modelling of the DSMC data. \\

We make analogous cell-centred assignments for the boundary and stress corrections at other faces, for both velocity ${\bf u}$ and scalar pressure $p$, and the stress correction tensor $\Phi$ appearing in the momentum equation \eqref{eq:NavierStokesMomentum}. 

\subsection{Surrogate models}

\label{ssec:surrogate-models}

To compute smooth corrections, for low-speed flows, we introduce surrogate models for each of the flow velocity ${\bf u}$, scalar pressure $p$ and the stress tensor $\bm{\tau}$. These are surrogates to the particle-based DSMC data and, by using sparse Bayesian learning, we obtain noise-reduced regression estimates for ${\bf u}$, $p$ and $\bm{\tau}$. \\

We consider the following set of variables containing the DSMC data: 
\begin{equation}
	\hat{\mathcal{D}} = \{ \hat{u},  \hat{v},  \hat{p},  \hat{\tau}_{xx},  \hat{\tau}_{yy},  \hat{\tau}_{xy} \}.
\end{equation}
Each variable in the particle-based dataset $\hat{\mathcal{D}}$ consists of two-dimensional cell-centred data evaluated on a discretised Cartesian grid $\mathcal{C}$:   
\begin{equation}
	\begin{split}
		\mathcal{C} = \left\{ (x_{i}, y_{j}) = \left(-L_{x}/2 + \left(i - 1/2 \right)\Delta x, -L_{y}/2 + \right. \right. \\ \left.\left.   \left(j - 1/2 \right)\Delta y\right) : i = 1 \rightarrow N_{x}, j = 1 \rightarrow N_{y} \right\},
	\end{split}
\end{equation}
where $\Delta x = L_{x} / N_{x}$ and $\Delta y = L_{y} / N_{y}$. Given the two-dimensional data $\hat{\bf{d}} \in \hat{\mathcal{D}}$ evaluated on $\mathcal{C}$, we transform to one-dimension via the bijection
\begin{equation}
	\hat{d}_{ij} \rightarrow \hat{d}_{k = i + (j-1)\times N_{x}}, \label{eq:bijection}
\end{equation}
where $i = 1 \rightarrow N_{x}, j = 1 \rightarrow N_{y}$. Following this bijection, we associate a surrogate model for each $\hat{{\bf d}} \in \mathcal{D}$ by decomposing the particle-based data into smooth and noisy parts:
\begin{equation}
	\hat{{\bf d}} = \bar{{\bf d}} + {\boldsymbol \epsilon}_{\bf d}, \label{eq:Decomposition}
\end{equation}
where $\bar{{\bf d}}$ is the smooth part and ${\boldsymbol \epsilon}_{\bf d}$ the noisy part. Our choice for $\bar{{\bf d}}$ is linear in some unknown weights ${\bf w}$ to be found:
\begin{equation}
	\bar{{d}}_{k} = (\Psi {\bf w})_{k} = \sum_{n=1}^{B} w_{n} \varphi_{n}({\bf x}_{k}),  \label{eq:Linearity}
\end{equation}
where $\Psi$ is the design matrix constructed using multilevel RBFs $\varphi_{n}({\bf x}_{k}), n = 1 \rightarrow B$ evaluated on a Cartesian grid point ${\bf x}_{k} = (x_{k},y_{k}) \in \mathcal{C}$ (see Section \ref{sec:multilevel-radial-basis-functions}). The unknown weights ${\bf w}$ are to be found using sparse Bayesian learning (see Section \ref{sec:sparse-bayesian-learning}). In addition, we assume uncorrelated, additive Gaussian noise:
\begin{equation}
	{\boldsymbol \epsilon}_{\bf d} \sim \mathcal{N}({\bf 0}, \beta^{-1}\mathds{1}), \label{eq:GaussianNoise}
\end{equation}
where the hyperparameter $\beta$ is the inverse-variance, or precision, of the Gaussian noise. This assumption, together with the assumption of independence between ${\bf w}$ and ${\boldsymbol \epsilon}_{\bf d}$, gives a multivariate Gaussian likelihood function with diagonal covariance: $\hat{{\bf d}} \vert {\bf w} \sim \mathcal{N}(\Psi {\bf w}, \beta^{-1}\mathds{1})$. The surrogate model associated to ${\hat {\bf d}} \in \hat{\mathcal{D}}$ can be concisely written as
\begin{equation}
	\hat{{\bf d}} = \Psi {\bf w} + {\bm \epsilon}_{\bf d}, \label{eq:TargetEquation}
\end{equation}
where $\hat{\bf d}$ is an one-dimensional vector of length $N_{x} \times N_{y}$ consisting of the target DSMC values and ${\boldsymbol \epsilon}_{\bf d}$ is the additive Gaussian noise given by equation \eqref{eq:GaussianNoise}. The design matrix $\Psi$, first introduced in equation \eqref{eq:Linearity}, is a two-dimensional matrix of size $(N_{x} \times N_{y}) \times B$ given by
\begin{equation}
	\Psi_{kn} = \varphi_{n}({\bf x}_{k}). \label{eq:DesignMatrix}
\end{equation}
Thus, $\Psi$ consists of $N_{x} \times N_{y}$ Cartesian grid points as the rows and $B$ radial basis functions as the columns. \\

From our choice of surrogate models, we see that they are linear in the weights ${\bf w}$ and the Gaussian noise ${\bm \epsilon}_{\bf d}$ is uncorrelated. This simplicity allows for analytical algorithms to be designed for inference. In particular, an uncorrelated Gaussian noise allows us to obtain analytical posterior distributions for the automated Bayesian learning.

\section{Sparse Bayesian learning to compute smooth corrections}

\label{sec:sparse-bayesian-learning}

\begin{figure}
	\centering
	\includegraphics[scale=0.32]{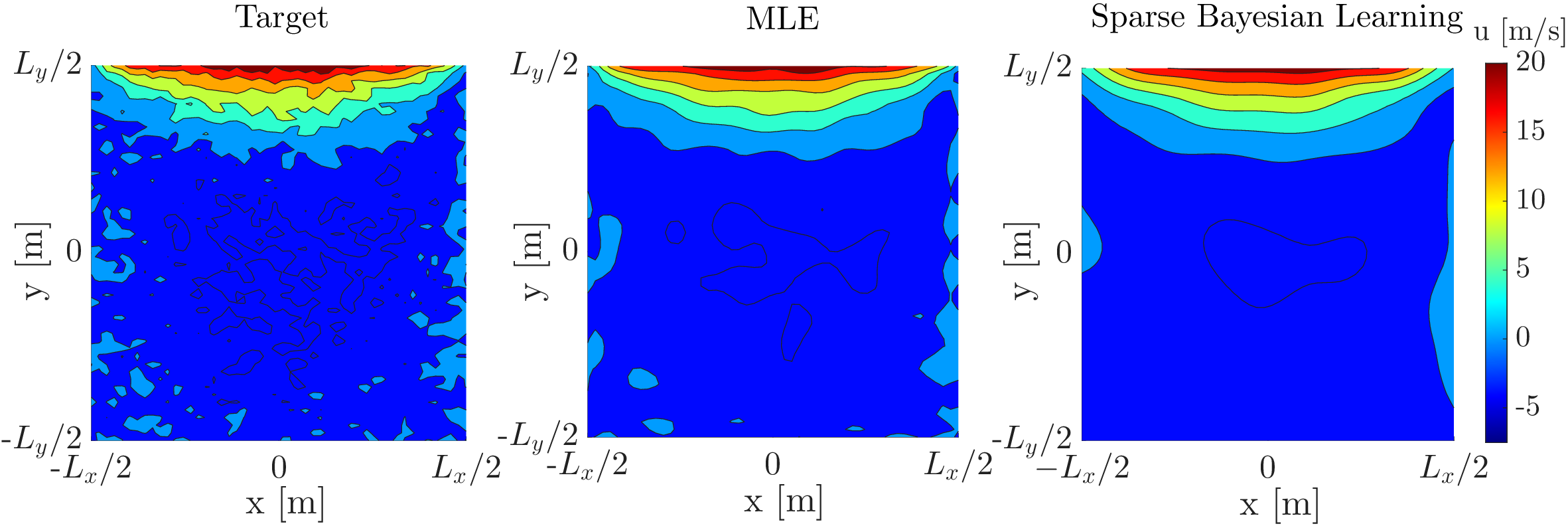}
	\caption{ A comparison between MLE and sparse Bayesian learning for noisy target DSMC data for the x-component of velocity, $u$. The regression estimates are found using the same design matrix $\Psi$. }
	\label{fig:MLE_vs_SparseBayesianLearning_Comparison}
\end{figure}

In order for our hybrid approach to be robust in reducing DSMC noise and computing smooth corrections, we use sparse Bayesian learning. In comparison to frequentist approaches such as the maximum likelihood estimate (MLE), Bayesian learning is less prone to overfitting. For example, Figure \ref{fig:MLE_vs_SparseBayesianLearning_Comparison} shows a comparison of regression estimates using MLE and sparse Bayesian learning for some noisy target DSMC data. Here, the same design matrix $\Psi$ is used to obtain the regression estimates from MLE and sparse Bayesian learning. We see clearly that sparse Bayesian learning retains less noise in its estimate compared to MLE. \\

As a result of equation \eqref{eq:TargetEquation}, the likelihood function, defined by the probability distribution of $p(\hat{{\bf d}} \vert {\bf w})$, is given by Gaussian distribution $\mathcal{N}(\Psi {\bf w}, \beta^{-1}\mathds{1})$. Using a Bayesian approach, we assume that the prior distribution for the weights ${\bf w}$ is also a Gaussian distribution: ${\bf w} \sim \mathcal{N}({\bf 0}, {\bf A}^{-1}), {\bf A} = \mathrm{diag}({\bm \alpha}) = \mathrm{diag}(\{ \alpha_{n}, n = 1 \ldots B \})$. This means that, before we see any data, we assume a Gaussian prior for the weights that has zero mean and diagonal covariance. Here, the hyperparameters ${\bm \alpha}$ are the inverse-variances, or precisions, associated to each of the weights. In particular, when a component of ${\bm \alpha}$ tends to infinity, the corresponding component of the weights ${\bf w}$ goes to zero (i.e. the matching basis vector is pruned from the basis set). 

\subsection{Posterior distribution}

Given this choice of Gaussian prior and Gaussian likelihood $\hat{{\bf d}} \vert {\bf w} \sim \mathcal{N}(\Psi {\bf w}, \beta^{-1}\mathds{1})$, we are now interested in calculating the posterior probability distribution of the weights ${\bf w}$ given the observed data $\hat{{\bf d}}$: $p({\bf w} \vert \hat{{\bf d}})$. It can be shown that the posterior distribution is also Gaussian given by $p({\bf w} \vert \hat{\bf d}) = \mathcal{N}({\bf m}, {\bm \Sigma})$ \cite{bishop2006pattern}, where the mean and covariance are given by
\begin{eqnarray}
	{\bf m} = \beta {\bm \Sigma} \Psi^{T} \hat{{\bf d}}, \label{eq:MeanSSBL} \\
	{\bm \Sigma} = ({\bf A} + \beta \Psi^{T} \Psi)^{-1}. \label{eq:CovarianceSSBL} 
\end{eqnarray}
The derivation of this result can be found in \cite{bishop2006pattern} (pages 90 - 92). We are interested in finding the optimal hyperparameters $\bm{\alpha}$ and $\beta$, appearing in equations \eqref{eq:MeanSSBL} and \eqref{eq:CovarianceSSBL}, that maximises the log marginal likelihood of the observed DSMC data:

\begin{equation}
	\ln p({\hat{\bf d}}\vert \bm{\alpha}, \beta)) = -\frac{1}{2} \left\{ N_{x}N_{y}\ln(2\pi) + \ln\vert {\bf C}\vert + \hat{{\bf d}}^{T} {\bf C}^{-1} \hat{{\bf d}} \right\},
\end{equation}

where ${\bf C} = \beta^{-1}\mathds{1} + \Psi {\bf A}^{-1} \Psi^{T}$ is an $N_{x}N_{y} \times N_{x}N_{y}$ matrix. To do this, we employ an existing fast sparse Bayesian learning algorithm \cite{tipping2003fast} based on the original relevance vector machine (RVM) \cite{tipping2001sparse} algorithm. 

\subsection{Sequential Sparse Bayesian Learning}

\begin{figure}
	\centering
	\includegraphics[scale=0.30]{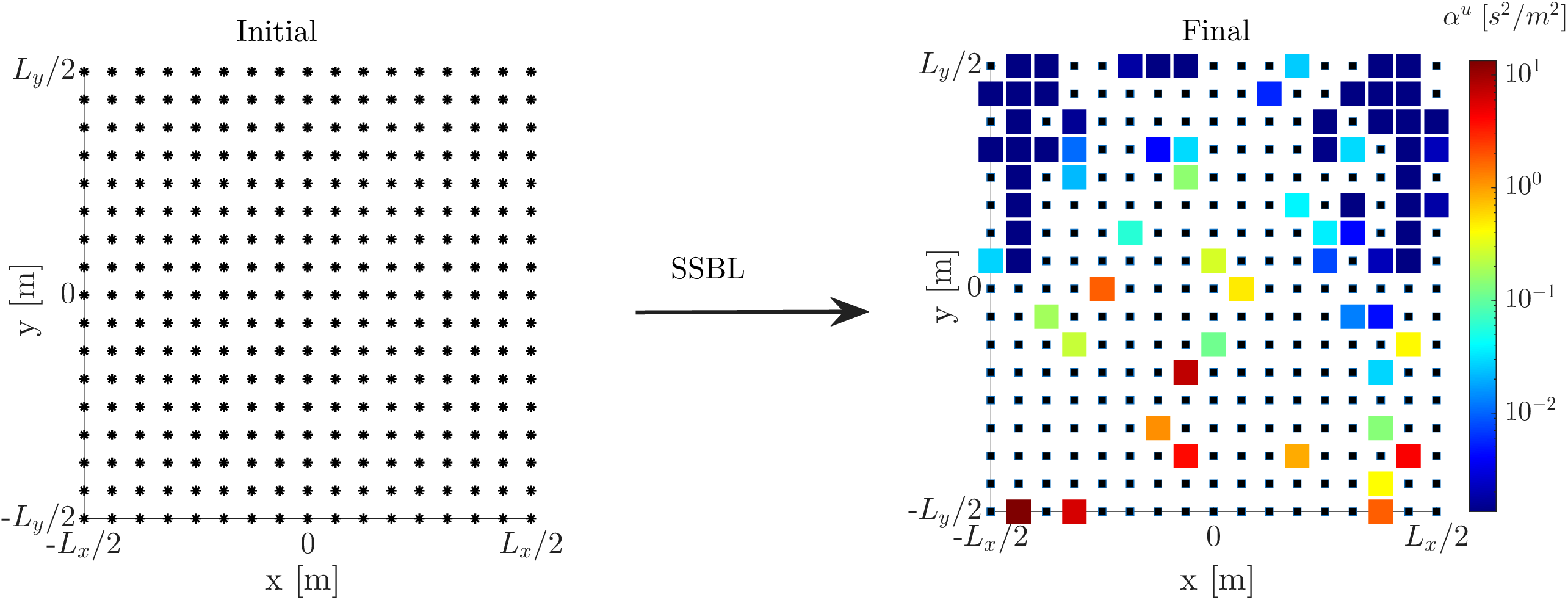}
	\caption{ An example of the application of the SSBL algorithm for the x-component of velocity, $u$, for a square cavity. Here, basis vectors are added based on `relevance', starting with an initial empty set. Finite values of the hyperparameters ${\bm \alpha}$ correspond to final basis functions in the model. }
	\label{fig:Alpha_Pruning_Diagram_Kn_0_05_M_0_1_0_5_Percent_Level4_2}
\end{figure}

The RVM algorithm was originally introduced in \cite{tipping2001sparse}, where regression on one- and two-dimensional data sets was performed with squared exponential kernels. In addition to flat, non-informative priors for the hyperparameters, derivations were also presented using more general gamma distributed priors. As well as presenting error improvements in employing RVM over other machine learning algorithms such as support vector machines (SVMs), they also explored, for one- and two-dimensional datasets, a qualitative comparison between target variables and regression estimates. \\

Since the original RVM algorithm starts with the entire design matrix before pruning, which can result in an initial high memory cost, a fast implementation was then introduced in \cite{tipping2003fast}, where one starts with an empty basis set and adds basis vectors by calculating its `relevance' (see Figure \ref{fig:Alpha_Pruning_Diagram_Kn_0_05_M_0_1_0_5_Percent_Level4_2} for an example in relation to a LDC test case). The analytical framework of this approach is detailed in \cite{faul2001analysis}. This algorithm is known as the `Sequential Sparse Bayesian Learning' (SSBL), and we use this fast algorithm to learn the hyperparameters $\bm{\alpha}$ and $\beta$ in the main results of this paper. The steps of this algorithm can be found in \cite{bishop2006pattern}, and in \cite{chinnappan2025bayesian} where it was applied for time-dependent flows. Since the hyperparameters are learned through this algorithm, the regression analysis is automatic, satisfying one of our original requirements for the proposed DSMC-CFD coupling. \\

Although the SSBL algorithm is fast and efficient, the algorithm, however, assumes that the prior distributions for $\bm{\alpha}$ and $\beta$ are flat, non-informative priors (i.e. distributions that are uniform on $\log(\alpha_{i})$ and $\log(\beta)$). In \ref{sec:gamma-hyperprior}, we revisit this assumption, providing results under more general gamma distributed hyperpriors using the original algorithm presented in \cite{tipping2001sparse}. This more general approach is important in suppressing the effects of overfitting when regression is performed on DSMC data with very high noise.

\subsubsection{Update approach}

Before proceeding, it should also be noted that simplicity of the RVM algorithm allows an `update approach' to be applied to achieve significant acceleration when the design matrix $\Psi$ and/or target vector ${\bf {\hat d}}$ may not fit in memory (see \cite{chinnappan_data_and_code} for further details). This is particularly useful for time-dependent problems, where one may have a long time series of $N_{T}$ steps. In the algorithms from \cite{tipping2001sparse} and \cite{tipping2003fast}, updates can be made noting that the linear models \eqref{eq:TargetEquation} at points $J$ and $J+1$ are related via the following:
\begin{equation}
	\tilde{{\bf d}} = \tilde{\Psi} {\bf w} + \tilde{\bm \epsilon},
\end{equation}
where

$ \tilde{\Psi} = \begin{bmatrix}
	\Psi_{1:J} \\ \Psi_{J+1}
\end{bmatrix} $, $\tilde{{\bf d}} = \begin{bmatrix}
	\hat{{\bf d}}_{1:J} \\ \hat{{\bf d}}_{J+1} \end{bmatrix}$ and $ \tilde{\bm \epsilon} = \begin{bmatrix}
	{\bm \epsilon}_{1:J} \\ {\bm \epsilon}_{J+1} \end{bmatrix}$. Therefore, we can update three quantities of importance as follows:
\begin{eqnarray}
	\tilde{\Psi}^{T}\tilde{\Psi} = \Psi^{T}_{1:J}\Psi_{1:J} + \Psi^{T}_{J+1}\Psi_{J+1}, \\
	\tilde{{\bf d}}^{T}\tilde{{\bf d}} = \hat{{\bf d}}_{1:J}^{T}\hat{{\bf d}}_{1:J} + \hat{{\bf d}}_{J+1}^{T}\hat{{\bf d}}_{J+1}, \\
	\tilde{\Psi}^{T}\tilde{{\bf d}} = \Psi^{T}_{1:J}\hat{{\bf d}}_{1:J} + \Psi^{T}_{J+1}\hat{{\bf d}}_{J+1}. 
\end{eqnarray}
From this observation, we see that a storage of only $B\times B$ for $\Psi^{T}\Psi$, $1 \times 1$ for $\tilde{{\bf d}}^{T}\tilde{{\bf d}}$ and $B \times 1$ for $\tilde{\Psi}^{T}\tilde{{\bf d}}$ is required when inferring the hyperparameters $\bm{\alpha}$ and $\beta$ for one-dimensional time-dependent flows. This is in contrast to $N_{T} \times B$ and $N_{T} \times 1$ needed for direct storage of $\Psi$ and $\hat{{\bf d}}$. 

\subsection{Corrections}

Once the optimal hyperparameters ${\bm \alpha}$ and $\beta$ have been found using the SSBL algorithm, we can use the mean of the posterior distribution ${\bf m}$, given by equation \eqref{eq:MeanSSBL}, for each surrogate model to estimate their unknown weights ${\bf w}$. Multiplying by the design matrix $\Psi$, we obtain the regression estimates from surrogate models: 
\begin{equation}
	\bar{{\bf d}} = \Psi {\bf m}.     
\end{equation}
Given this one-dimensional vector, we can apply the inverse of the map \eqref{eq:bijection} to obtain the two-dimensional estimate. The stress corrections tensor $\Phi$ is the difference between the inferred stress from the micro-model and the Newtonian constitutive stress obtained from the velocity gradients, i.e., $\tau_{ij} = -\mu\left(\frac{\partial u_{i}}{\partial x_{j}} + \frac{\partial u_{j}}{\partial x_{i}}\right)$. Therefore, we can compute $\Phi$ appearing in the N-S momentum equations \eqref{eq:NavierStokesMomentum} analytically using 
\begin{equation}
	\phi_{ij} = \bar{\tau}_{ij} + \mu \left( \frac{\partial \bar{u}_{i}}{\partial x_{j}} +  \frac{\partial \bar{u}_{j}}{\partial x_{i}} \right), \label{eq:Stress-Corrections}
\end{equation}
where $\mu = \rho \nu$ is the dynamic viscosity, and $\rho$ is the mass density that is calculated from input to the microscopic model. The velocity and pressure boundary corrections are obtained by evaluating $\bar{u}, \bar{v}$ and $\bar{p}$ at the four walls $\partial \Omega$.  In addition, before use in the icoFoam solver, the stress corrections $\Phi$ and scalar pressure $p$ are scaled by the mass density $\rho$ as in equation \eqref{eq:NavierStokesMomentum}.

\subsubsection{Reduced viscosity via minimisation}

	At higher $Kn$, we minimise the stress correction tensor $\Phi$ for $\mu$ as follows:
	\begin{equation}
		\mu^{\star} = \arg \min_{\mu > 0} \left[ \int_{-L_{y}/2}^{L_{y}/2} \int_{-L_{x}/2}^{L_{x}/2} \vert\vert \Phi \vert\vert_{F}^{2} dx dy \right],  \label{eq:ReducedViscosityMinimisation}
	\end{equation}
	where we have used the Frobenius norm denoted by $F$. For \(Kn=0.05\) and \(0.1\), the physical dynamic viscosity of argon is used. For $Kn > 0.1$, where non-equilibrium effects are expected to dominate, the Newtonian stress term is large relative to the inferred stress term, due to the assumed constant viscosity in this work. Therefore, we introduce a viscosity minimisation to moderate the Newtonian contribution, enabling the surrogate model to more robustly capture the non-Newtonian component of the stress correction tensor. This effective viscosity parameter $\mu^{\star}$, or precisely the kinematic version $\nu^{\star} = \mu^{\star}/ \rho$, is then used in the continuum-based, icoFoam solver to obtain the CFD estimates. Therefore, the DSMC corrections and the modified momentum equations are written as: 

	\begin{eqnarray}
		\phi_{ij} = \bar{\tau}_{ij} + \mu^{\star} \left( \frac{\partial \bar{u}_{i}}{\partial x_{j}} +  \frac{\partial \bar{u}_{j}}{\partial x_{i}} \right), \label{eq:Stress-Corrections-withviscosityminisation}  \\
		\frac{\partial {\bf u}}{\partial t} + ({\bf u} \cdot \nabla) {\bf u} - \nabla \cdot (\nu^{\star} \nabla {\bf u}) + \frac{1}{\rho} \nabla \cdot \Phi = -\frac{1}{\rho} \nabla p, \label{eq:NavierStokesMomentumwithviscosityminimisation}
	\end{eqnarray}

It is important to note that the effective viscosity parameters, \(\mu^\star\) and \(\nu^\star\), are introduced as parameters of the DSMC-informed stress decomposition and are not intended to represent the physical viscosity or kinematic viscosity of the working gas.

\subsubsection{Credible intervals}

In addition to regression estimates, we can also compute credible intervals of $95\%$ for the macroscopic quantities of interest. To do this, for each of the components of velocity, scalar pressure and stress tensor, we generate a sample from their posterior distributions (i.e. $\mathcal{N}({\bf m}, {\bm \Sigma})$ with \eqref{eq:MeanSSBL} and \eqref{eq:CovarianceSSBL}). We then multiply by the design matrix $\Psi$ to obtain estimates based on this sample. Evaluating at the four walls for velocity and scalar pressure gives the boundary corrections. By using expression \eqref{eq:Stress-Corrections}, we obtain the stress corrections, where we replace the bar variables with that generated from the sample. These are then used in the icoFoam solver for the N-S equations \eqref{eq:NavierStokesMomentum} - \eqref{eq:NavierStokesContinuity} to obtain a CFD sample for the velocity, scalar pressure and stress. Performing this procedure for $N_{s} = 50$ many realisations gives us a distribution of values for velocity, scalar pressure and stress. Using the empirical cumulative distribution function, we can identify the $2.5\%$ and $97.5\%$ percentiles to obtain a $95\%$ credible interval that is based on the uncertainty of the weights ${\bf w}$. 

\section{Multilevel radial basis functions to obtain regression estimates}

\label{sec:multilevel-radial-basis-functions}

While MMS-Sparse in Chinnappan et al. \cite{chinnappan2025bayesian} used a complete basis set of Chebyshev polynomials for the spatial dimension, for more complex geometries in two- and three-dimensions, RBFs can be more flexible, particularly when the geometry is unbounded. For example, in thin extended regions to a surface, Chebyshev polynomials cannot easily be applied in an automated way. We thus propose multilevel RBFs that can be automated to learn both broad and fine details of noisy, spatially two-dimensional datasets. By using a multilevel approach, we can methodologically set the level of detail to retain in the regression analysis of the noisy DSMC datasets. While we do not demonstrate our multilevel RBFs approach beyond the cavity problem, it can in principle be applied to more complex two-dimensional geometrical configurations than considered in this paper, such as unbounded flow past an arbitrary solid body. In such an example, the RBFs can be defined over an equidistant set of points over the physical domain of interest, where RBFs inside the solid body (i.e. where there is no flow defined) are removed from consideration.  This, therefore, provides a potential route to treat more complex configurations. \\

\begin{figure}
	\centering
	\includegraphics[scale=0.35]{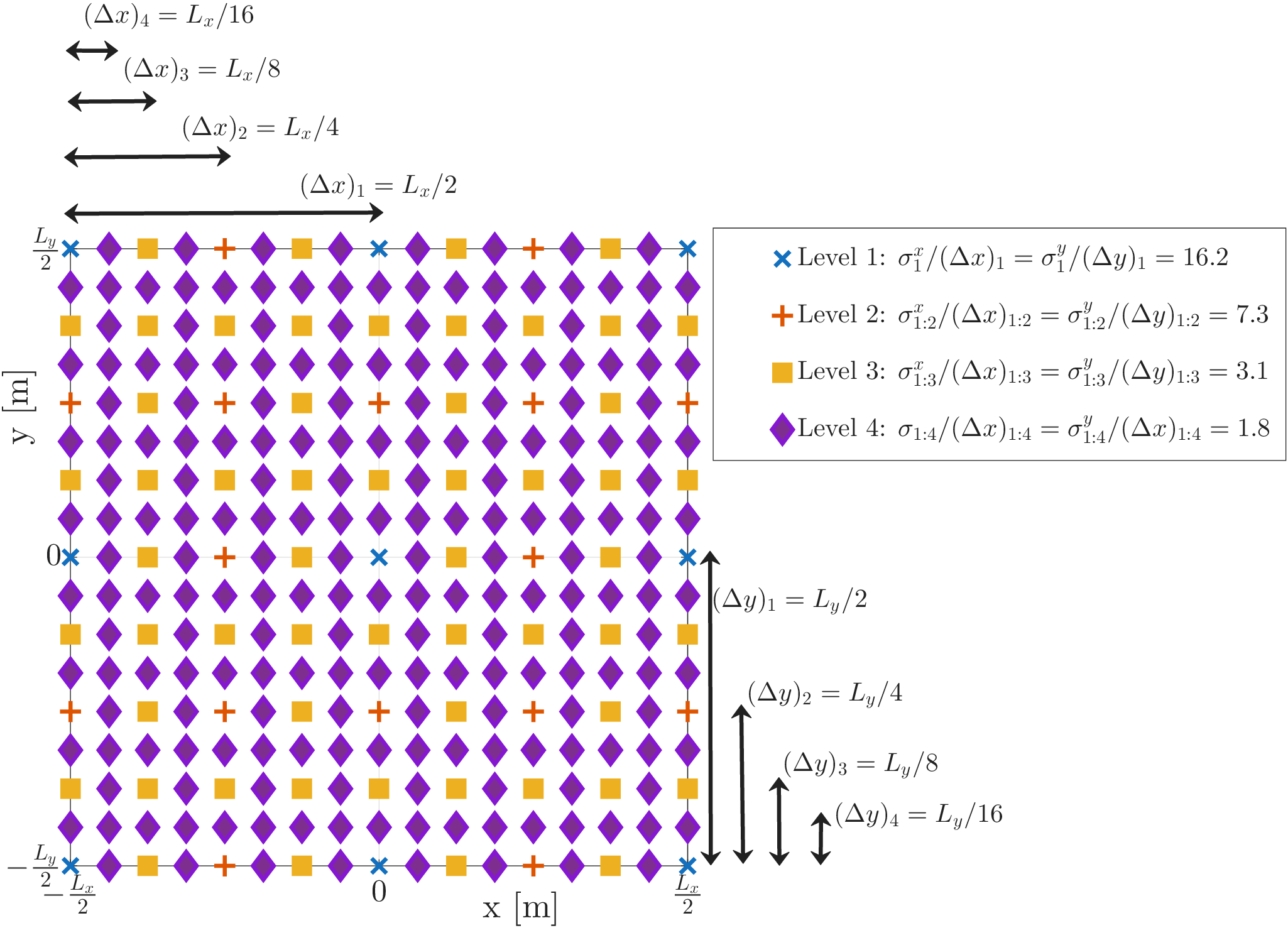}
	\caption{ Centres of the multilevel RBFs up to level 4. Each symbol represents a different level. The value $\sigma_{1:l}^{x} / (\Delta x)_{1:l} = \sigma_{1:l}^{y} / (\Delta y)_{1:l} $ is constant for RBFs in a given level $l$ and is chosen based on the condition number of design matrix $\Psi$. }
	\label{fig:Multilevel_RBFs_Diagram}
\end{figure}

We use squared exponential functions, defined on an equidistant two-dimensional grid, for the RBFs \cite{fasshauer2007meshfree} \cite{abu2012learning}:
\begin{equation}
	\varphi_{n}({\bf x}_{k}) = \exp\left(- \frac{\left( x_{k} - x_{n} \right)^{2}}{2(\sigma^{x}_{n})^{2}} - \frac{\left( y_{k} - y_{n} \right)^{2}}{2(\sigma^{y}_{n})^{2}} \right), \label{eq:RBFs}
\end{equation}
where ${\bf x}_{k} = (x_{k}, y_{k})$ is the $k$'th Cartesian grid point, and ${\bf x}_{n} = (x_{n}, y_{n})$, $\sigma^{x}_{n}$ $\&$ $\sigma^{y}_{n}$ are centre-points and the standard deviations of the $n$'th RBF respectively. In choosing the centres of the RBFs, we decompose an equidistant grid into levels, where higher levels correspond to an increased resolution, and RBFs in a given level are assigned the same scaled standard deviations.  \\

Each level $l$ has $B = (2^{l} + 1) \times (2^{l} + 1)$-many RBFs with centres defined by 
\begin{equation}
	(x_{n}, y_{m}) = \left( - \frac{L_{x}}{2} + \frac{L_{x}}{2^{l}} \cdot (n-1), - \frac{L_{y}}{2} + \frac{L_{y}}{2^{l}} \cdot (m-1) \right), \label{eq:RBFsCentres}
\end{equation}
for $n = 1 \rightarrow 2^{l}+1$ and $m = 1 \rightarrow 2^{l}+1$.  Therefore, there are $3\times 3$ RBFs at level 1, $5\times 5$ at level 2, $9\times 9$ at level 3, $17\times 17$ at level 4, $33\times 33$ at level 5, $65\times 65$ at level 6, and so on. \\

A diagram on the centre locations and standard deviations up to and including level 4 is shown in Figure \ref{fig:Multilevel_RBFs_Diagram}.  In each level $l$, we set the same scaled standard deviation to the RBFs: $\sigma^{x}_{1:l} / (\Delta x)_{1:l} = \sigma^{y}_{1:l} / (\Delta y)_{1:l} = \kappa$. The value of $\kappa$ is chosen to be the smallest number, to one decimal point, that provides a reciprocal condition number for the design matrix $\Psi$ greater than the tolerance $> 10^{-12}$. Between successive levels, any centre locations that overlap with a lower level are assigned scaled standard deviations associated to the higher level. \\

When deciding which level to fix, we found that using RBFs up to and including level 4 can adequately capture the necessary required features of the low-noise benchmark DSMC data (see \ref{sec:RBFs-level} for more details). Conversely, for high-noise data, level 4 was seen to filter out, in many cases, the unnecessary higher-order features not ideal as input to the macro-CFD model.

\section{Proof-of-concept of MMS-Sparse} 
\label{sec:proof-of-concept}

We start by presenting the proof-of-concept of MMS-Sparse hybrid, where we use particle-based, DSMC data for the entire domain as input (see Figure \ref{fig:MMS-Sparse_OneWayCoupling}a). 

\subsection{Simulation parameters}

\begin{table}[]
	\begin{center} 
		\begin{tabular}{|l|l|l|l|l|l|l|}
			\hline
			\textbf{AR} & \textbf{M} & \textbf{Kn} & \textbf{$n_{\rho}$ [1/m$^{3}$] } & \textbf{N$_{x}$} & \textbf{N$_{y}$} & \textbf{$\nu$ / $\nu^\star$ [m$^{2}$/s]} \\ \hline
			1           & 0.1        & 0.05        & 2.59E19                    & 50          & 50          & 12.3                               \\ \hline
			1           & 0.1        & 0.1         & 1.29E19                    & 50          & 50          & 24.7                               \\ \hline
			1           & 0.1        & 0.5         & 2.59E18                    & 50          & 50          & 6.85$^{\star}$          \\ \hline
			1           & 0.1        & 5           & 2.59E17                    & 50          & 50          & 7.81$^{\star}$          \\ \hline
			0.5         & 0.2        & 0.5         & 2.59E18                    & 100         & 50          & 8.11$^{\star}$           \\ \hline
			2           & 0.2        & 0.5         & 2.59E18                    & 50          & 100         & 8.85$^{\star}$           \\ \hline
		\end{tabular} 
	\end{center}
	\caption{ Table of DSMC parameters using SPARTA \cite{plimpton2019direct} and CFD parameters using icoFoam \cite{greenshields2021}. For \(Kn=0.05\) and \(0.1\), \(\nu\) denotes the physical kinematic viscosity of argon. For \(Kn>0.1\), \(\nu^\star\) denotes the effective kinematic viscosity parameter obtained from the stress-correction minimisation in equation \ref{eq:ReducedViscosityMinimisation}.}
	\label{table:Simulation_Parameters}
\end{table}

We consider both square and rectangular domains given by aspect ratio $AR := L_{y} / L_{x} = 1$ and $AR = 0.5$ $\&$ $2$ respectively. We consider argon as the gas species. For a square cavity with $AR = 1$, we consider Knudsen numbers $Kn = \lambda / L = 0.05, 0.1, 0.5$ $\&$ $5$ at Mach number $M = U_{w} / U_{a} = 0.1$, where $U_{a} = 308$ m/s is the speed of sound in the argon gas species at $273$ K. For rectangular cavities $AR = 0.5$ $\&$ $2$, we consider Knudsen number $Kn = 0.5$ at $M = 0.2$. For both square and rectangular cavities, we choose the characteristic length to be the side length $L = \min\{L_{x}, L_{y}\} = 1$ m. To compute the mean free path $\lambda$, we use the following expression \cite{bird1994molecular}:
\begin{equation}
	\lambda = \left( \sqrt{2}\pi D_{ref}^{2} n_{\rho} (T_{ref}/T)^{\omega - 1/2} \right)^{-1}, \label{eq:MeanFreePathBirdsEquation}
\end{equation}
where $D_{ref} = 4.17 \times 10^{-10}$ m, $n_{\rho}$ is the number density, $T_{ref} = T = 273$ K, and $\omega = 0.81$ for VHS argon gas. \\

A summary of the different simulation parameters considered in the running of the DSMC and CFD is shown in Table \ref{table:Simulation_Parameters}. For the DSMC, we initialise $N_{P} = 2.5 \times 10^{7}$ test particles for the square cavity cases and $N_{P} = 5 \times 10^{7}$ test particles for the rectangular cavity cases. This corresponds to a particle per cell, $PPC = 10000$ in our simulations. While using a fewer number of particles per cell would decrease the computational cost, it would also increase the level of DSMC noise. The solid surfaces are assumed to undergo diffuse reflection with full thermal accommodation at a prescribed wall temperature of $273$ K.  For all the test cases considered, the time-step for the DSMC simulations performed using SPARTA \cite{SPARTA-DSMC} was chosen to be $\Delta t = 1\times 10^{-7}$ s, while we used $N_{steady} = 8 \times 10^{5}$ steps to reach steady-state in the system at time $T_{steady} = 0.08$ s. After steady-state, we perform data collection every $500$ steps for a further $3 \times 10^{5}$ steps. And so, each DSMC simulation was completed at $t = 0.11$ s. We refer to the time-averaged DSMC data over $0.03$ seconds after steady-state as the benchmark DSMC. For the training DSMC, we average up to a tenth of this duration (i.e. up to and including $3 \times 10^{4}$ steps after steady-state). We note that the time to reach steady state is larger than the time for the averaging. This is because a high number of particles per cell ($PPC = 10000$) reduces statistical noise, allowing a shorter sampling window to obtain benchmark results. A lower PPC would require significantly longer averaging to achieve comparable noise levels, which could lead to a sampling duration that exceeds the steady-state duration. For the CFD using icoFoam \cite{greenshields2021}, the resolution is the same as that used for the DSMC. The time-step $\Delta t$ and the end times of the CFD simulations were set to be $1\times 10^{-4}$ s and $6$ s respectively. In particular, we found that the end time of $6$ s was sufficient to establish steady-state, where the number of iterations needed to solve for the velocity and scalar pressure is zero. In addition to MMS-Sparse (i.e. CFD with corrections), we also performed Pure CFD simulations, i.e. CFD with no corrections, at $AR = 1, 0.5$ $\&$ $2$, with Mach numbers $M = 0.1$ $\&$ $0.2$. Here, the kinematic viscosity was set to be equal to that used for MMS-Sparse. \\ 

For the simulation parameters considered, generally speaking, we found that off-diagonal shear stress $\tau_{xy}$ was most dominant in magnitude in the stress tensor ${\bm \tau}$. And so, estimates for diagonal components, $\tau_{xx}$ $\&$ $\tau_{yy}$ are not shown in the results of this paper. 

\subsection{Aspect ratio, $AR = 1$}

\subsubsection{Slip regime}

\begin{figure}
	\centering
	\includegraphics[scale=0.40]{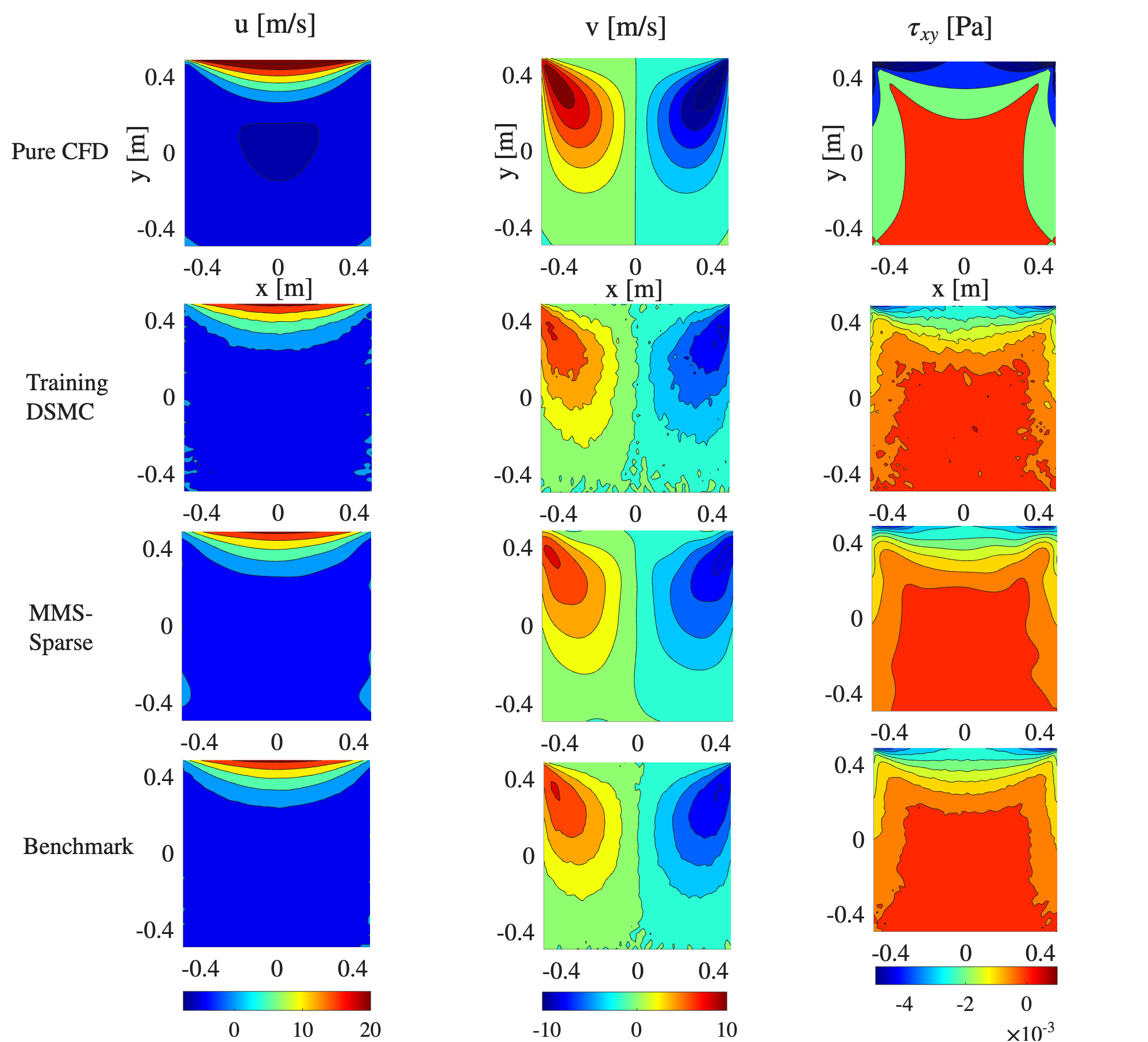}
	\caption{ Comparison between Pure CFD, training DSMC, MMS-Sparse and benchmark DSMC at $Kn = 0.05$ $\&$ $M = 0.1$. }
	\label{fig:JointPlot_MMS_Hybrid_10Percent_Kn_0_05_M_0_1_Whole_Domain}
\end{figure}

We start with results in the slip regime $Kn = 0.05$ for a square cavity $AR = 1$ and Mach number $M = 0.1$. Figure \ref{fig:JointPlot_MMS_Hybrid_10Percent_Kn_0_05_M_0_1_Whole_Domain} shows the comparison between Pure CFD, training DSMC, MMS-Sparse, and benchmark DSMC (i.e. DSMC calculation with 10 times greater averaging) for the macroscopic quantities $u, v$ and $\tau_{xy}$. Our results, as well as showing good agreement between the MMS-Sparse hybrid and the benchmark DSMC, highlights the improvements that can be obtained when smooth corrections are supplied to a CFD solver. We also observe MMS-Sparse provides noise-reduced estimates for the macroscopic quantities. While the MMS-Sparse estimates are generally good compared to the benchmark, we note slight oscillations are present at the vertical side walls for $u$. \\

\begin{figure}
	\includegraphics[scale=0.40]{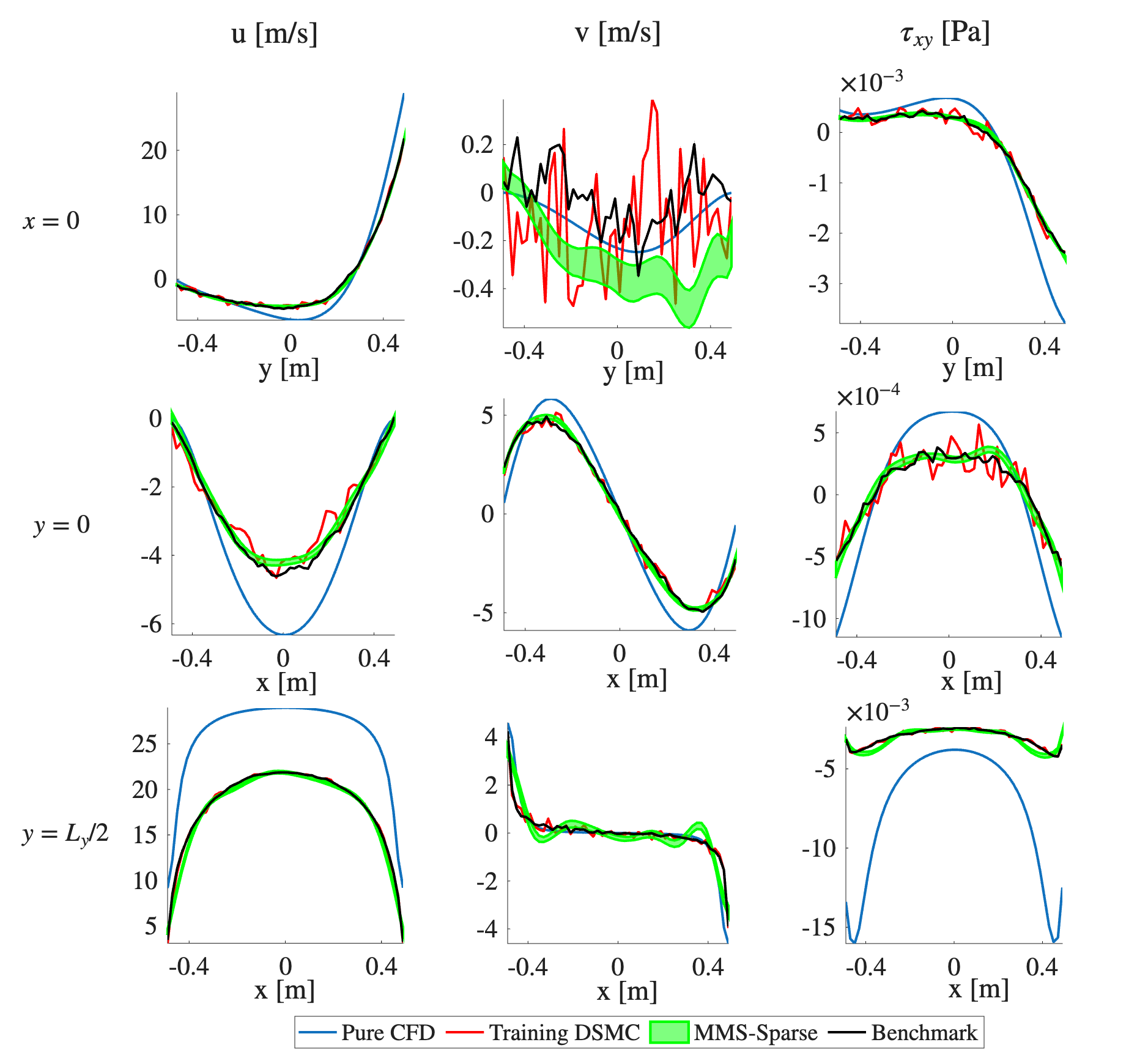}
	\centering 
	\caption{ Comparison at the centrelines and upper wall between Pure CFD, training DSMC, MMS-Sparse and benchmark DSMC at $Kn = 0.05$ $\&$ $M = 0.1$.  }	
	\label{fig:JointPlot_MMS_Hybrid_10Percent_Kn_0_05_M_0_1_2_WholeDomain_CredibleIntervals}
\end{figure}

In addition to the two-dimensional contour plots, we also present $95\%$ credible intervals for the three macroscopic quantities at the horizontal and vertical centrelines and at the upper wall. These are shown in Figure \ref{fig:JointPlot_MMS_Hybrid_10Percent_Kn_0_05_M_0_1_2_WholeDomain_CredibleIntervals}. As with the contour plots, we note considerable improvements can be made by supplying corrections to the CFD solver (i.e. when compared to Pure CFD). The estimates with MMS-Sparse are generally good, although deviations are noted at the vertical centreline for $v$ compared to the benchmark. Furthermore, we note that in some regions the benchmark results are outside the $95\%$ credible bounds. 

\subsubsection{Transitional regime}

We next test the MMS-Sparse hybrid at higher Knudsen numbers in the transitional regime, keeping in mind that the kinematic viscosity $\nu$ for the CFD simulations is now calculated through minimisation, as in equation \eqref{eq:ReducedViscosityMinimisation}. \\

\begin{figure}
	\centering
	\includegraphics[scale=0.35]{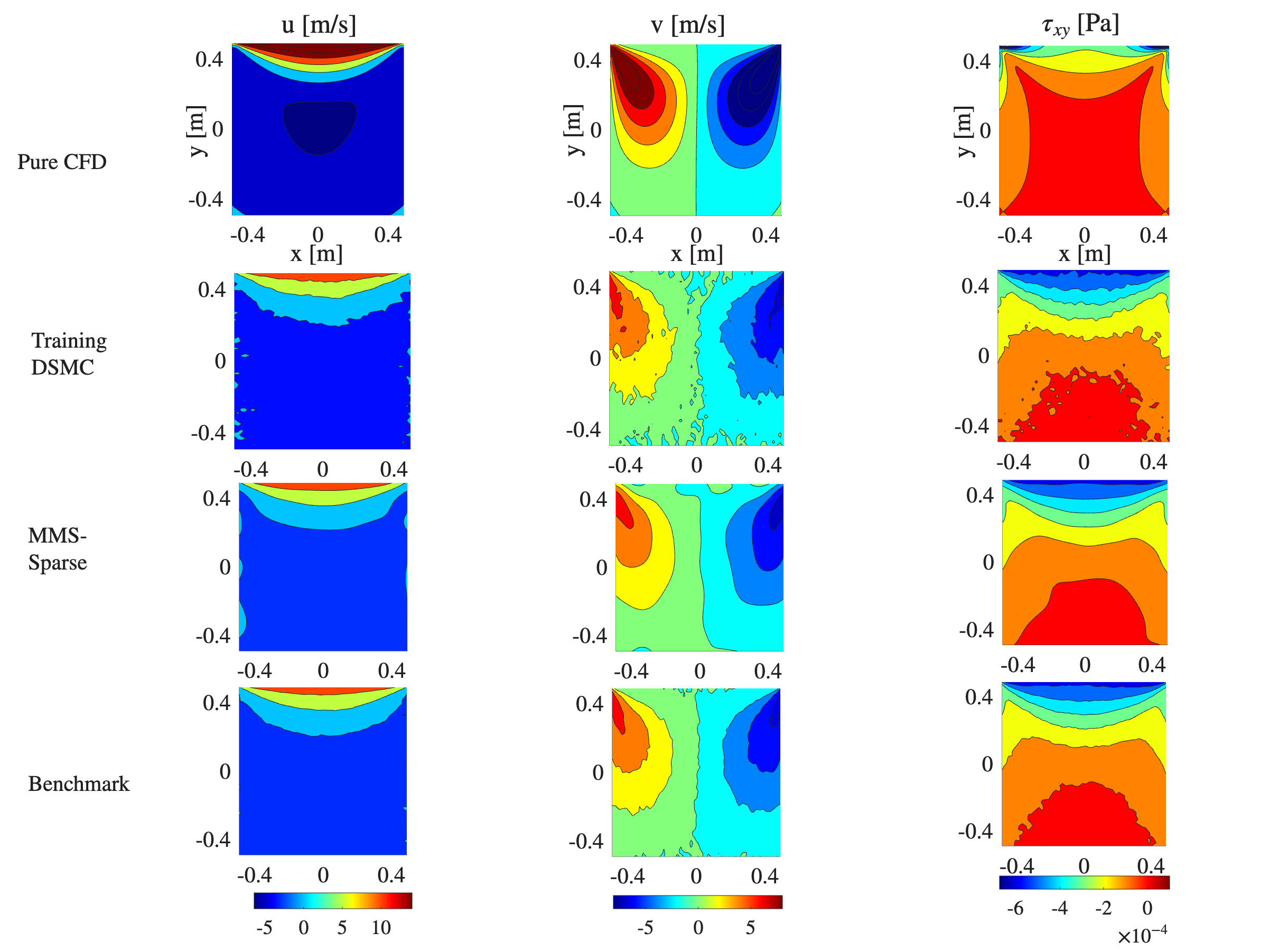}
	\caption{ Comparison between Pure CFD, training DSMC, MMS-Sparse and benchmark DSMC at $Kn = 0.5$ $\&$ $M = 0.1$. }
	\label{fig:JointPlot_MMS_Hybrid_10Percent_Kn_0_5_M_0_1}
\end{figure}

At $Kn = 0.5$ and $M = 0.1$, a comparison between Pure CFD, training DSMC, MMS-Sparse and benchmark DSMC is shown in Figure \ref{fig:JointPlot_MMS_Hybrid_10Percent_Kn_0_5_M_0_1}. As in the slip regime, we find good agreement from our estimates with MMS-Sparse, with improvements found compared to Pure CFD. Although slight oscillations are once again present at the side walls for $u$, we observe that the shear stress estimates are very good. \\

\begin{figure}
	\includegraphics[scale=0.35]{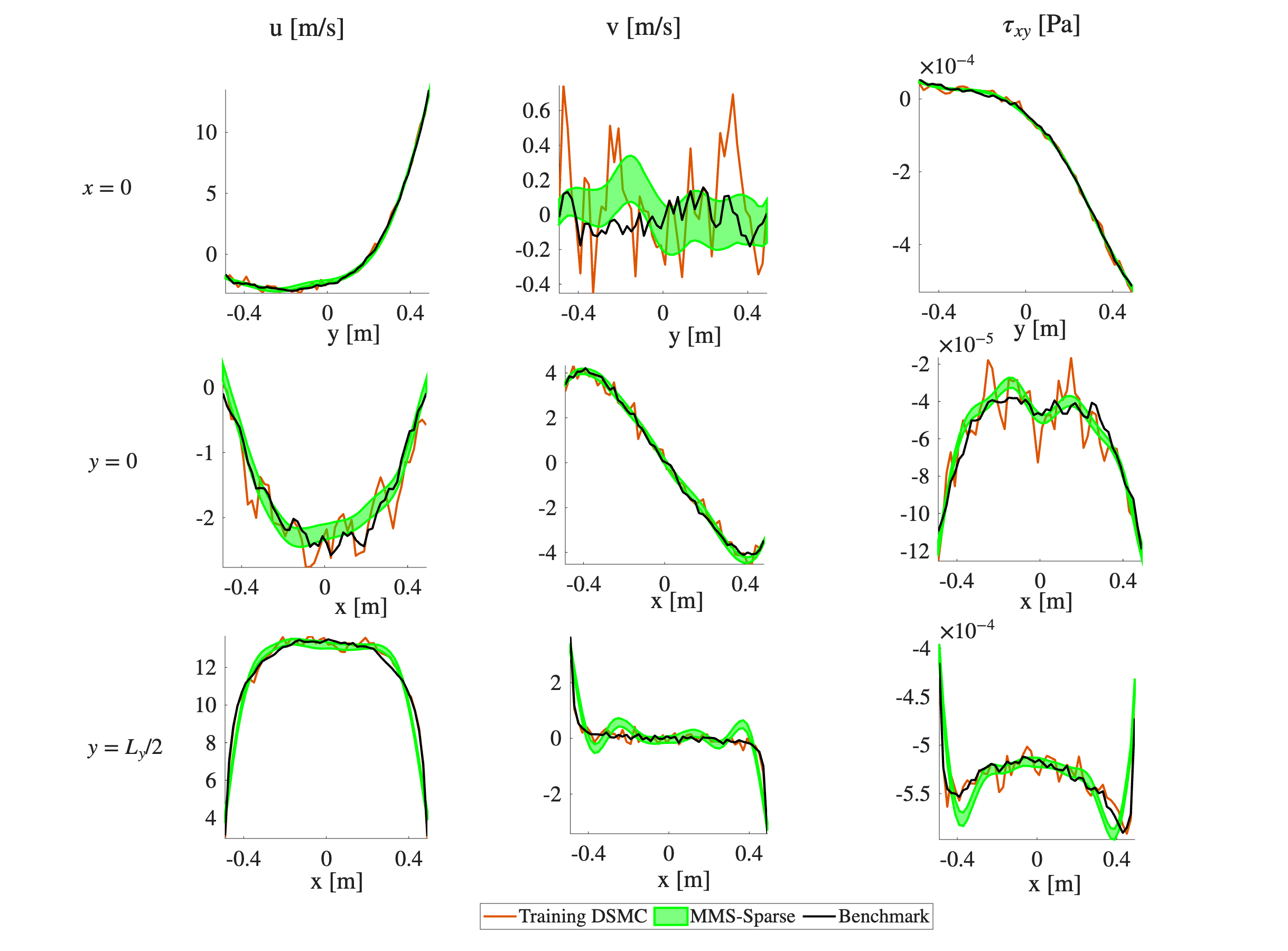}
	\centering 
	\caption{ Comparison at the centrelines and upper wall between training DSMC, MMS-Sparse and benchmark DSMC at $Kn = 0.5$ $\&$ $M = 0.1$.  }
	\label{fig:JointPlot_MMS_Hybrid_10Percent_Kn_0_5_M_0_1_WholeDomain_CredibleIntervals}
\end{figure}

The corresponding $95 \%$ credible intervals at the upper wall and centreline positions are shown in Figure \ref{fig:JointPlot_MMS_Hybrid_10Percent_Kn_0_5_M_0_1_WholeDomain_CredibleIntervals}. As in the slip regime, we find that MMS-Sparse provides generally good results compared to the benchmark DSMC. \\

\begin{figure}
	\centering
	\includegraphics[scale=.35]{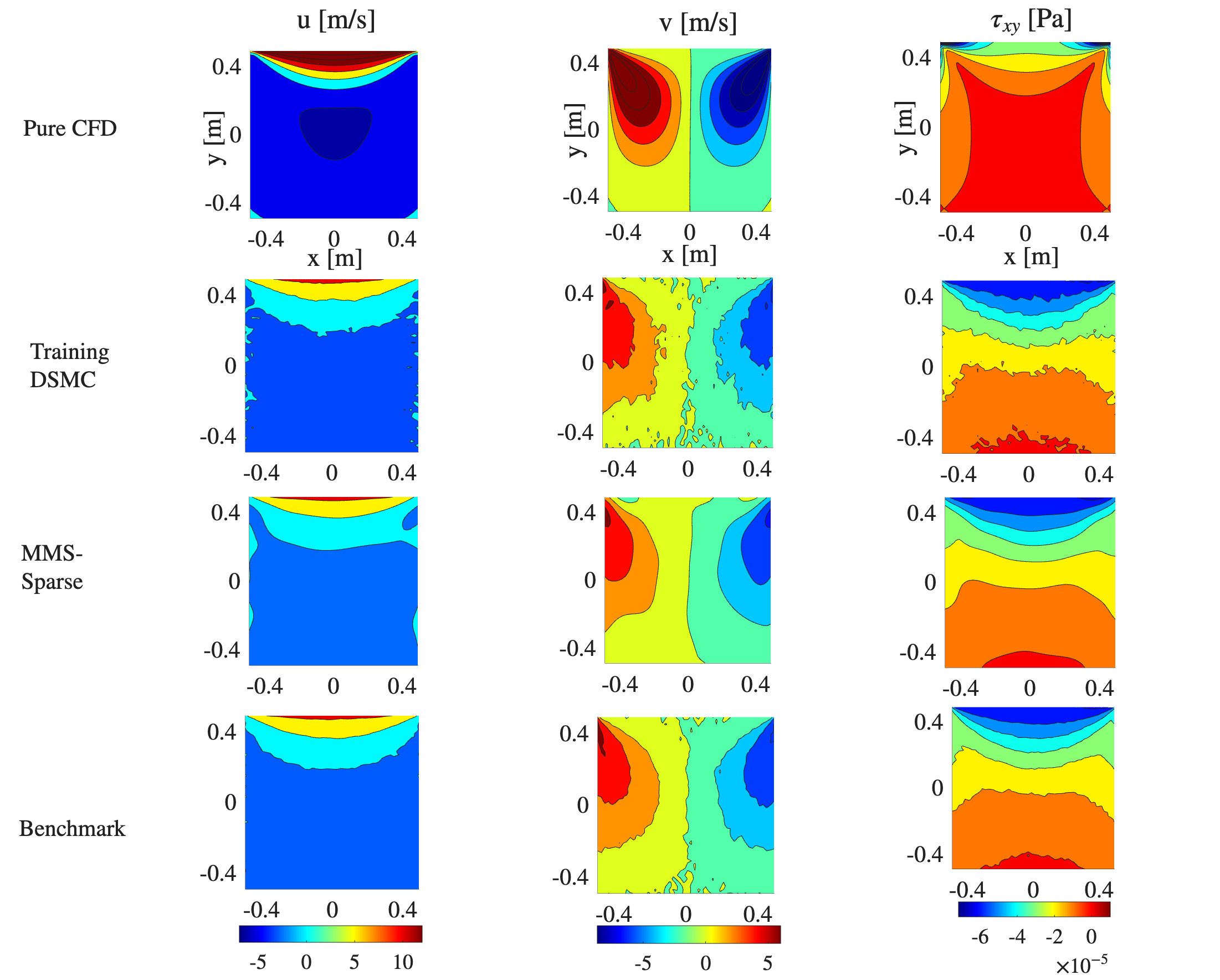}
	\caption{ Comparison between Pure CFD, training DSMC, MMS-Sparse and benchmark DSMC at $Kn = 5$ $\&$ $M = 0.1$.  }
	\label{fig:JointPlot_MMS_Hybrid_10Percent_Kn_5_M_0_1}
\end{figure}

The comparison for $Kn = 5$ is shown in Figure \ref{fig:JointPlot_MMS_Hybrid_10Percent_Kn_5_M_0_1}. Compared to the Pure CFD, we find considerable improvement with MMS-Sparse, when the training DSMC is used to compute the macroscopic quantities. However, we note that, for $u$, the agreement of MMS-Sparse with benchmark DSMC is less than the comparisons previously made at $Kn = 0.05$ $\&$ $0.5$. For example, the interior field for $u$ in the upper portion of the domain has clear deviations from the benchmark DSMC. Nevertheless, as with the other cases, the results for shear stress $\tau_{xy}$ are very good. \\

\begin{figure}
	\includegraphics[scale=.35]{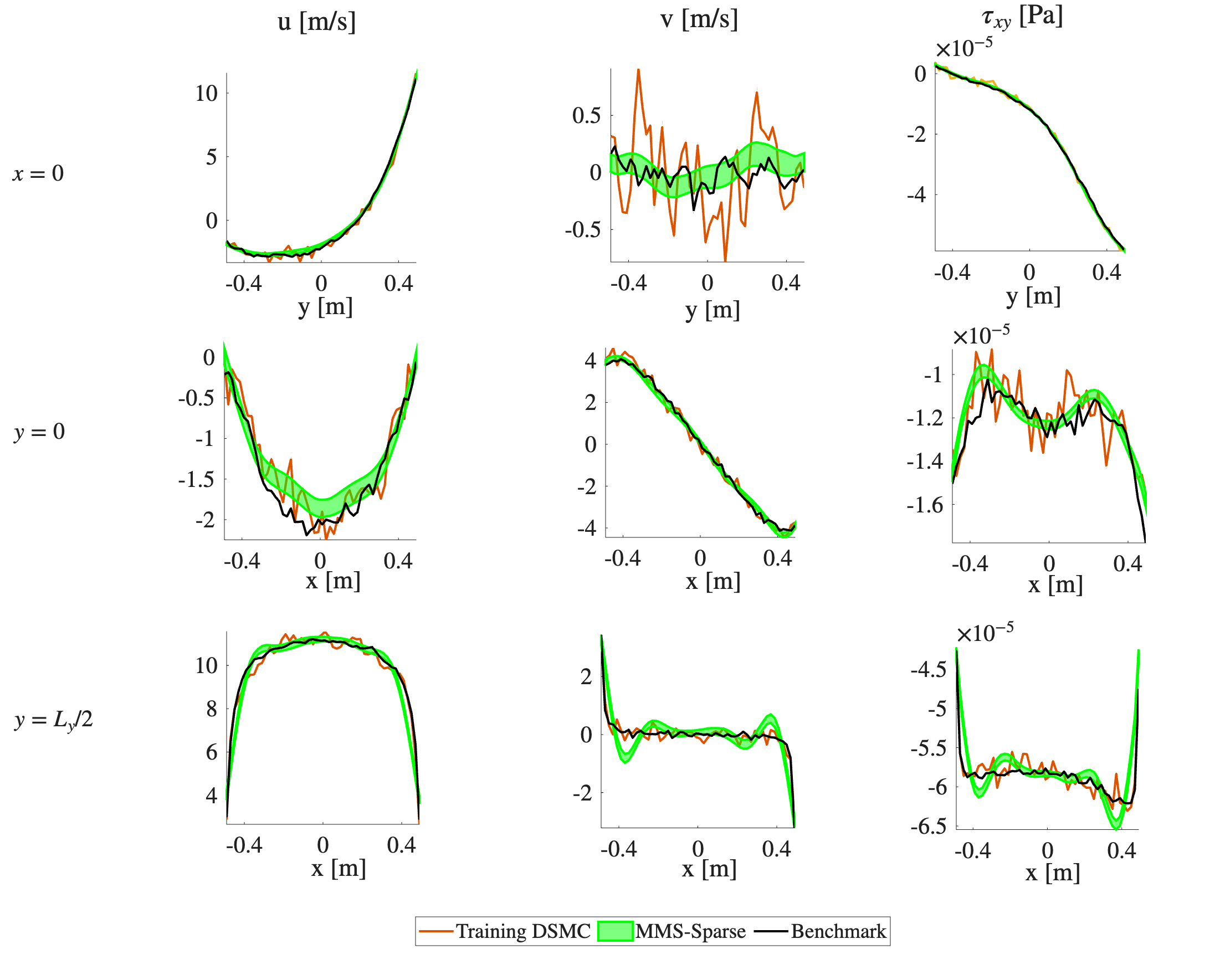}
	\centering 
	\caption{ Comparison at the centrelines and upper wall between training DSMC, MMS-Sparse and benchmark DSMC at $Kn = 5$ $\&$ $M = 0.1$. }
	\label{fig:JointPlot_MMS_Hybrid_10Percent_Kn_5_M_0_1_WholeDomain_CredibleIntervals}
\end{figure}

The $95 \%$ credible intervals are shown in Figure \ref{fig:JointPlot_MMS_Hybrid_10Percent_Kn_5_M_0_1_WholeDomain_CredibleIntervals}. At this higher $Kn = 5$, the results are, in general, good, even for $v$ at the vertical centreline. 

\subsection{Aspect ratios, $AR = 0.5$ $\&$ $2$}

We now proceed to consider, in the transitional regime $Kn = 0.5$, wide and tall cavities given by $AR = 0.5$ $\&$ $2$, with a higher Mach number $M = 0.2$. \\

\begin{figure}
	\centering
	\includegraphics[scale=0.35]{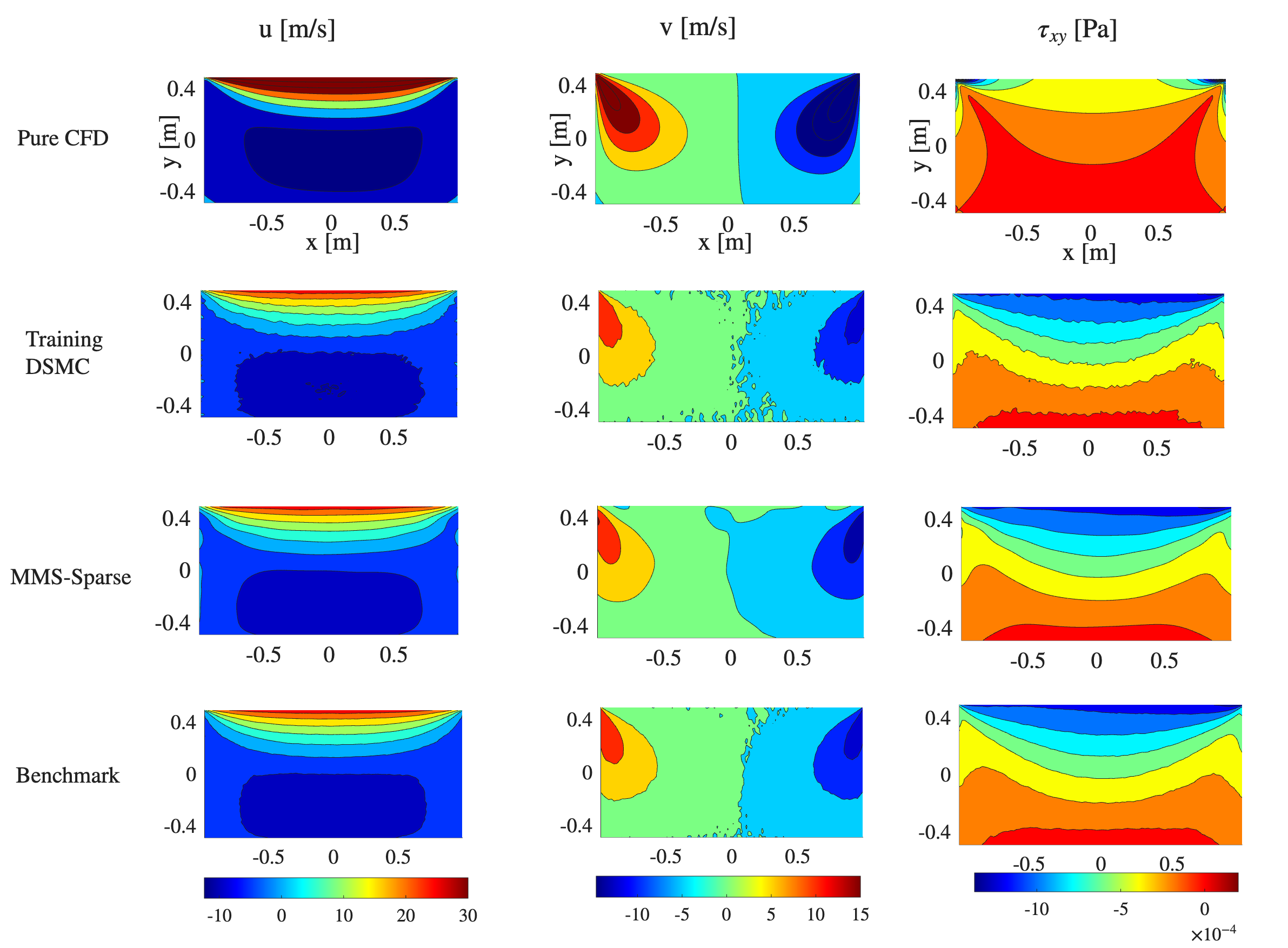}
	\caption{ Comparison between Pure CFD, training DSMC, MMS-Sparse and benchmark DSMC at $AR = 0.5, Kn = 0.5$ $\&$ $M = 0.2$.  }
	\label{fig:JointPlot_MMS_Hybrid_10Percent_AR_0_5_Kn_0_5_M_0_2}
\end{figure}

Starting at $AR = 0.5$ corresponding to a wide cavity, the comparison for the flow velocity and shear stress is shown in Figure \ref{fig:JointPlot_MMS_Hybrid_10Percent_AR_0_5_Kn_0_5_M_0_2}. Our results show that MMS-Sparse provides noise-reduced estimates that are good compared to benchmark DSMC. As with the square cavity cases, we note slight oscillations were again found at the side walls for $u$. Despite this, compared to the Pure CFD, we find considerable improvements with MMS-Sparse, when corrections are supplied to the CFD solver. \\

\begin{figure}
	\includegraphics[scale=0.35]{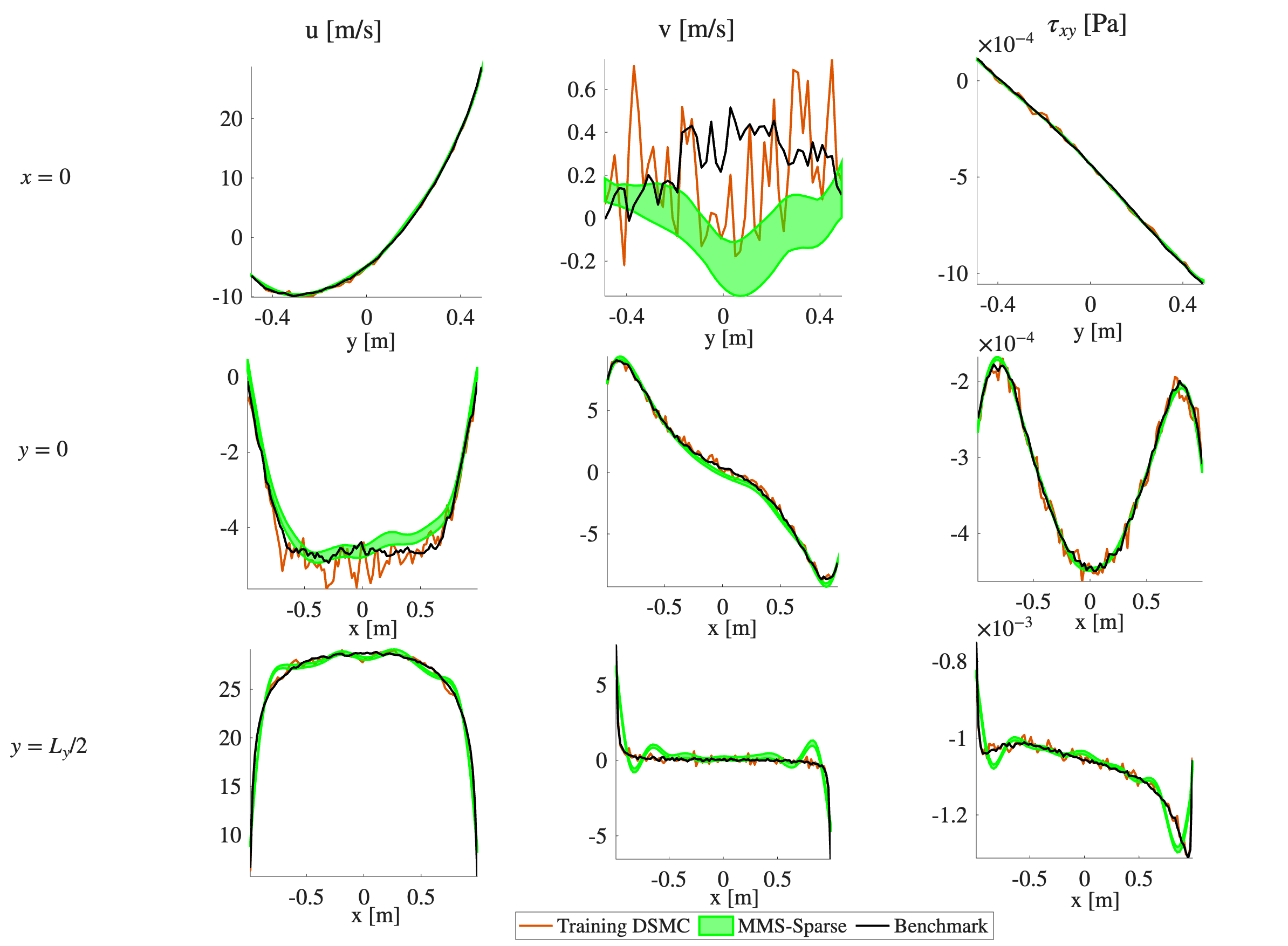}
	\centering 
	\caption{ Comparison at the centrelines and upper wall between training DSMC, MMS-Sparse and benchmark at $AR = 0.5, Kn = 0.5$ $\&$ $M = 0.2$. }
	\label{fig:JointPlot_MMS_Hybrid_10Percent_AR_0_5_Kn_0_5_M_0_2_WholeDomain_CredibleIntervals}
\end{figure}

Figure \ref{fig:JointPlot_MMS_Hybrid_10Percent_AR_0_5_Kn_0_5_M_0_2_WholeDomain_CredibleIntervals} shows the $95 \%$ credible intervals for this case. As with the square cavity cases, the estimates with MMS-Sparse are generally good, except at the vertical centreline for $v$ where deviations are noted. \\

\begin{figure}
	\centering
	\includegraphics[scale=0.35]{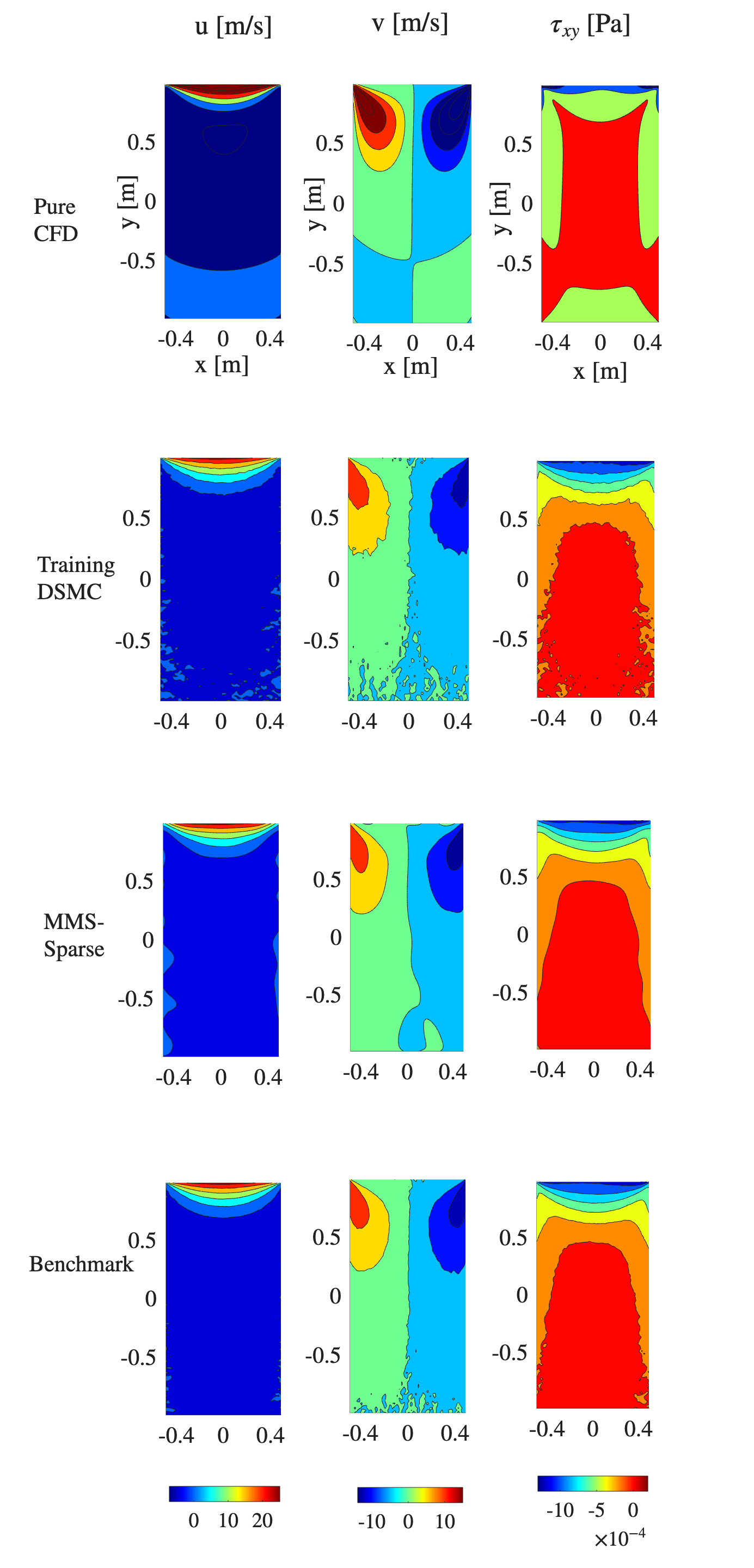}
	\caption{ Comparison between Pure CFD, training DSMC, MMS-Sparse and benchmark DSMC at $AR = 2, Kn = 0.5$ $\&$ $M = 0.2$. }
	\label{fig:JointPlot_MMS_Hybrid_10Percent_AR_2_Kn_0_5_M_0_2}
\end{figure}

We next proceed to examine a tall cavity $AR = 2$. Figure \ref{fig:JointPlot_MMS_Hybrid_10Percent_AR_2_Kn_0_5_M_0_2} shows the comparison. We find good agreement between MMS-Sparse and the benchmark DSMC, although oscillations are present at the side walls for $u$ and lower wall for $v$. Nevertheless, compared to Pure CFD, we find considerably improved estimates with MMS-Sparse, particularly for the shear stress. \\

\begin{figure}
	\includegraphics[scale=0.35]{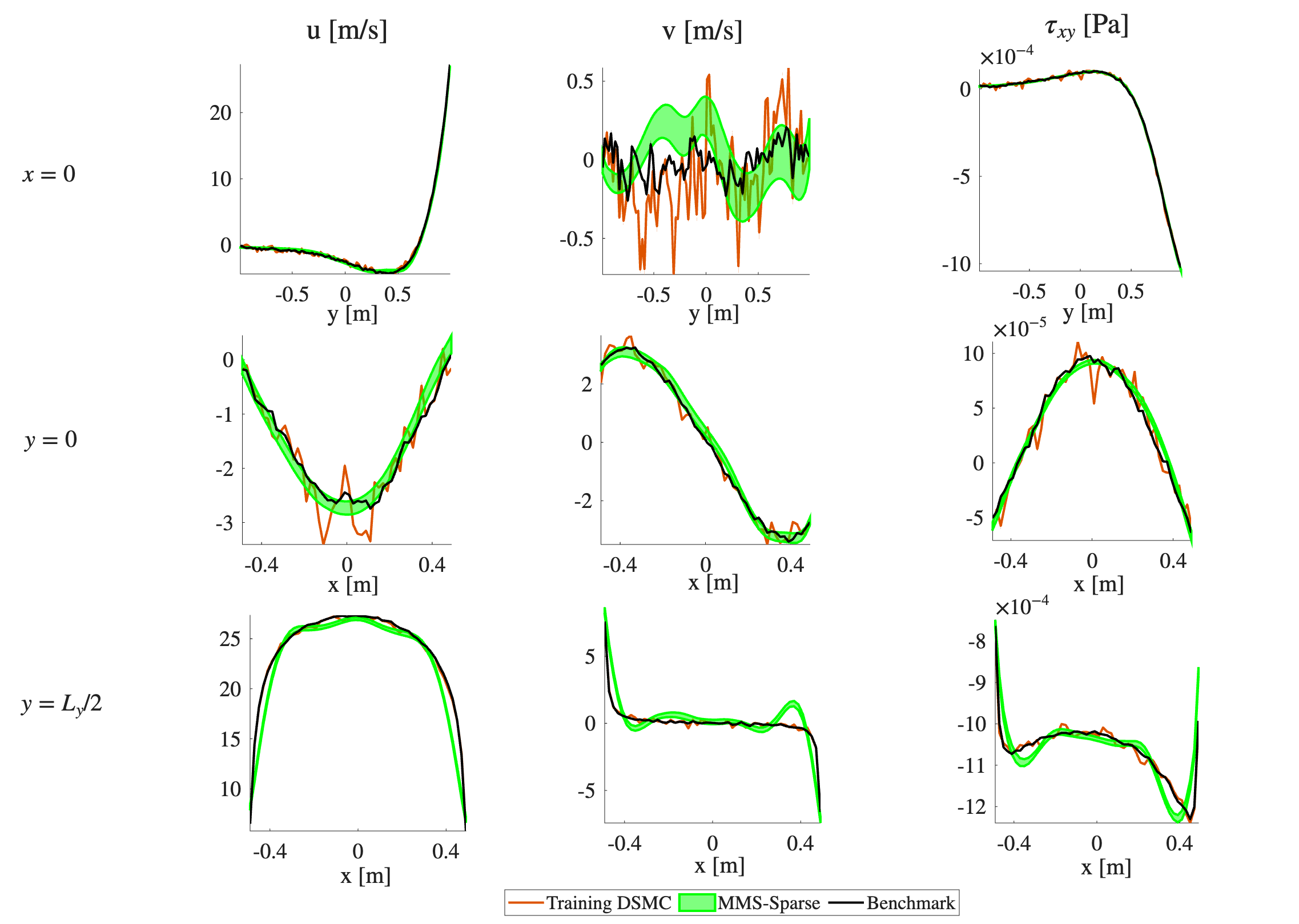}
	\centering 
	\caption{ Comparison at the centrelines and upper wall between training DSMC, MMS-Sparse and benchmark at $AR = 2, Kn = 0.5$ $\&$ $M = 0.2$. }
	\label{fig:JointPlot_MMS_Hybrid_10Percent_AR_2_Kn_0_5_M_0_2_WholeDomain_CredibleIntervals}
\end{figure}

The $95 \%$ credible intervals for the tall cavity case are shown in Figure \ref{fig:JointPlot_MMS_Hybrid_10Percent_AR_2_Kn_0_5_M_0_2_WholeDomain_CredibleIntervals}. Once more, we find good results with MMS-Sparse compared to the benchmark, although some deviations are again found at the vertical centreline for $v$. 

\subsection{In-principle computational savings}

\begin{table}[]
	\begin{center} 
		\begin{tabular}{|l|l|l|l|l|l|}
			\hline
			\textbf{AR} & \textbf{M} & \textbf{Kn} & \textbf{ Training  } & \textbf{ Inference } & \textbf{Benchmark  } \\ 
			&  &  & \textbf{ DSMC [mins] } & \textbf{ [mins] } & \textbf{ DSMC [mins] } \\ \hline
			1           & 0.1        & 0.05        &  657.6                    &   76.6     &     1174.1                           \\ \hline
			1           & 0.1        & 0.5         &   664.3                 &    69.5       &      1182.5      \\ \hline
			0.5         & 0.2        & 0.5         &     1758.6             &     57.6       &      2934.6     \\ \hline
			2           & 0.2        & 0.5         &    1700.6                 &     55.4      &      2875.6     \\ \hline
		\end{tabular}  
	\end{center}
	\caption{ Table of elapsed times (in minutes) for the training DSMC, inference (regression + CFD with corrections) and benchmark DSMC. }
	\label{table:Computational_Cost}
\end{table}

For our simulations, we also measured the elapsed times for the training DSMC, inference (regression + CFD with corrections) and benchmark DSMC. These are shown in Table \ref{table:Computational_Cost}. These timings demonstrate the in-principle computational savings found by using full-domain MMS-Sparse (training DSMC + inference) compared to the full-domain benchmark DSMC. For the four cases considered, we observe a speed up factor of approximately $1.6$ when using full-domain MMS-Sparse compared to full-domain benchmark DSMC. \\

The in-principle computational saving is limited, since the initial transient period required to reach a steady state, $N_{steady} = 8 \times 10^{5}$ timesteps is common to both the training and benchmark DSMC simulations. Thus, the above-mentioned computational saving is solely due to reducing the subsequent averaging period. Larger computational savings could be achieved by applying the present methodology within a true DSMC-CFD coupled framework, in which DSMC is initially restricted to localised non-equilibrium regions, and CFD is employed elsewhere. Once the DSMC data have been used to train the surrogate, the DSMC in the non-equilibrium regions can be replaced by the CFD with corrections. However, such savings remain potential and would need to be quantified through a fully coupled implementation with spatially restricted DSMC, which is left for future work.

\section{Pseudo-hybrid approach}
\label{sec:pseudo-hybrid-approach} 

While we have investigated the proof-of-concept of MMS-Sparse using DSMC data from the entire domain, in a hybrid framework, one may typically perform particle-based simulation only close to the boundaries, where the flow can be in non-equilibrium. To test MMS-Sparse in such a hybrid context, we use a `pseudo-hybrid approach', previously presented in \cite{chinnappan2025bayesian}. In this approach, we use DSMC data two mean free paths from the walls, i.e. $2\lambda$, consistent with the treatment adopted in our previous work \cite{chinnappan2025bayesian} and assign zero corrections interior in our input in MMS-Sparse (see Figure \ref{fig:MMS-Sparse_OneWayCoupling}b). The cut-off of $2\lambda$ is chosen since, at low-speeds, non-equilibrium effects near solid surfaces are confined within Knudsen layers, which are typically reported to extend approximately one to two mean free paths from the bounding surface \cite{karniadakis2006microflows}\cite{lockerby2005capturing}. This fixed wall-normal distance is adopted as a practical cut-off criterion for the present pseudo-hybrid demonstration, although a true hybrid formulation would require a more rigorous criterion for identifying the non-equilibrium regions, such as a local gradient-length Knudsen number \cite{vasileiadis2024hybriddcfoam}. Moreover, for the cases considered in the slip regime, the flow away from the walls is expected to be reasonably well described by the conventional N-S equations, with the stress corrections comparatively small in those regions. For future work, an adaptive criterion for identifying non-equilibrium regions, such as a gradient-based Knudsen number, could be adopted, as is commonly considered in DSMC-CFD coupling frameworks \cite{vasileiadis2024hybriddcfoam}. In this pseudo-hybrid demonstration, we only use multilevel RBFs that are within the $2\lambda$ distance from the walls to learn the constitutive corrections. In particular, before use in the SSBL algorithm, the design matrix is constructed with only RBFs a distance $\leq 2\lambda$ from the walls. And so, the regression analysis using the SSBL algorithm is still automated.  \\

This approach is intended primarily to demonstrate the capability of the MMS-Sparse model to reconstruct the required non-equilibrium corrections using DSMC data restricted to the near-wall region ($2\lambda$). This setup represents a simplified approximation of a true DSMC-CFD coupled framework, in which DSMC would be employed only in localised non-equilibrium regions while the remainder of the domain is treated using CFD.  Since the present pseudo-hybrid calculation still performs DSMC over the full domain, it is not intended to represent the computational cost of such a spatially restricted hybrid implementation.

\subsection{Slip regime}

\begin{figure}
	\centering
	\includegraphics[scale=0.35]{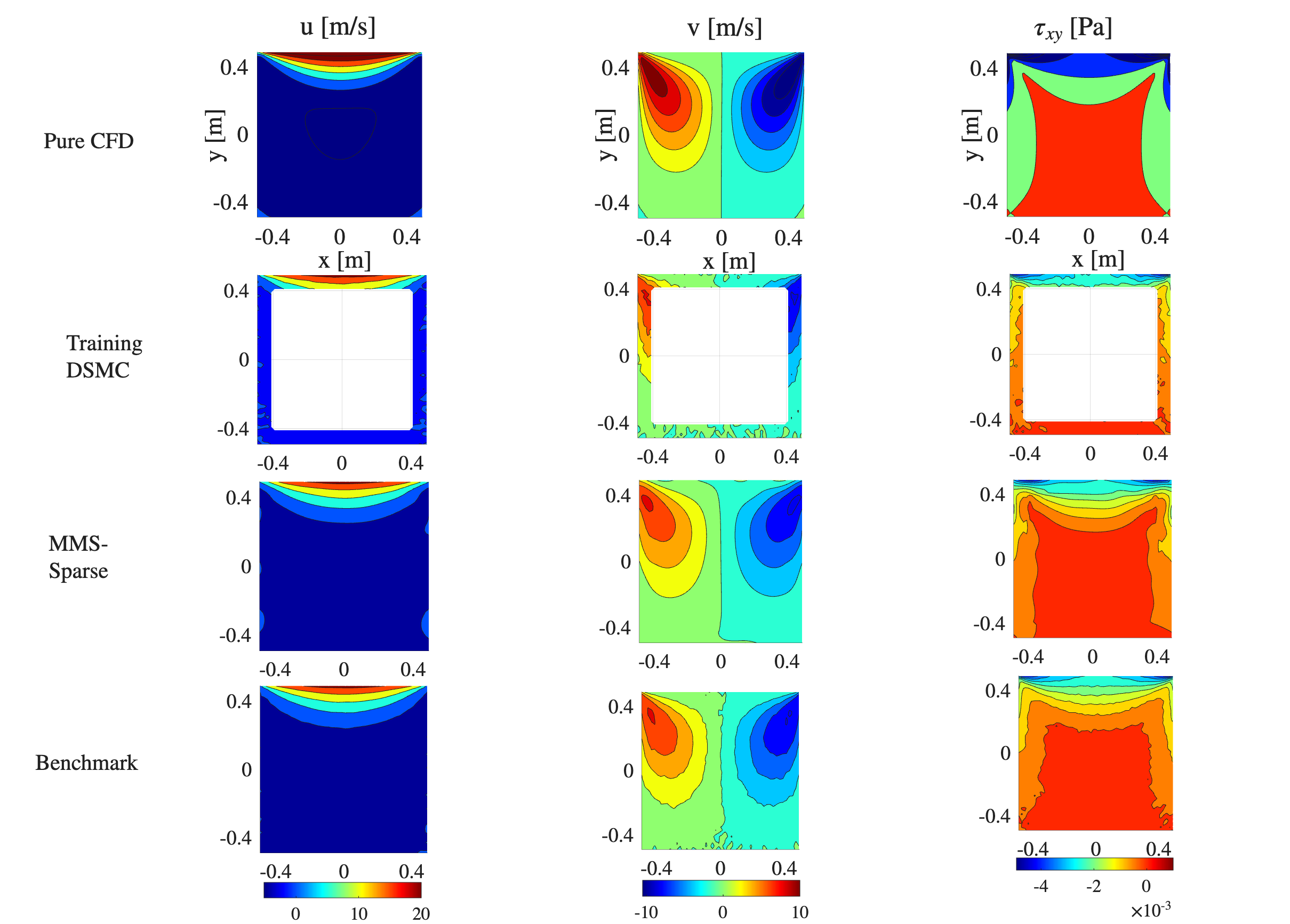}
	\caption{ Comparison between Pure CFD, training DSMC, MMS-Sparse with pseudo-hybrid approach, and benchmark DSMC at $Kn = 0.05$ $\&$ $M = 0.1$. }
	\label{fig:JointPlot_MMS_Hybrid_2Lambda_10Percent_Kn_0_05_M_0_1}
\end{figure}

We start at the slip regime $Kn = 0.05$ with $M = 0.1$. Figure \ref{fig:JointPlot_MMS_Hybrid_2Lambda_10Percent_Kn_0_05_M_0_1} shows the resulting comparison. Here, the second row shows the input DSMC, where only data $2\lambda = 0.1$ m is used as input. The third row shows the estimates from the pseudo-hybrid approach. Compared to the benchmark in the fourth row, we see that there is very good agreement for the velocity, while, for the shear stress, noticeable deviations are present towards the upper corner points of the cavity. This observation suggests that, for this specific LDC test case, although the pseudo-hybrid approach gives good estimates for flow velocity, there are noticeable deviations present for the shear stress. We note that such deviations were not present in the proof-of-concept (see Figure \ref{fig:JointPlot_MMS_Hybrid_10Percent_Kn_0_05_M_0_1_Whole_Domain}). Despite this, compared to the Pure CFD, we see that improvements can be found with MMS-Sparse when corrections are supplied to the CFD. \\

\begin{figure}
	\includegraphics[scale=0.35]{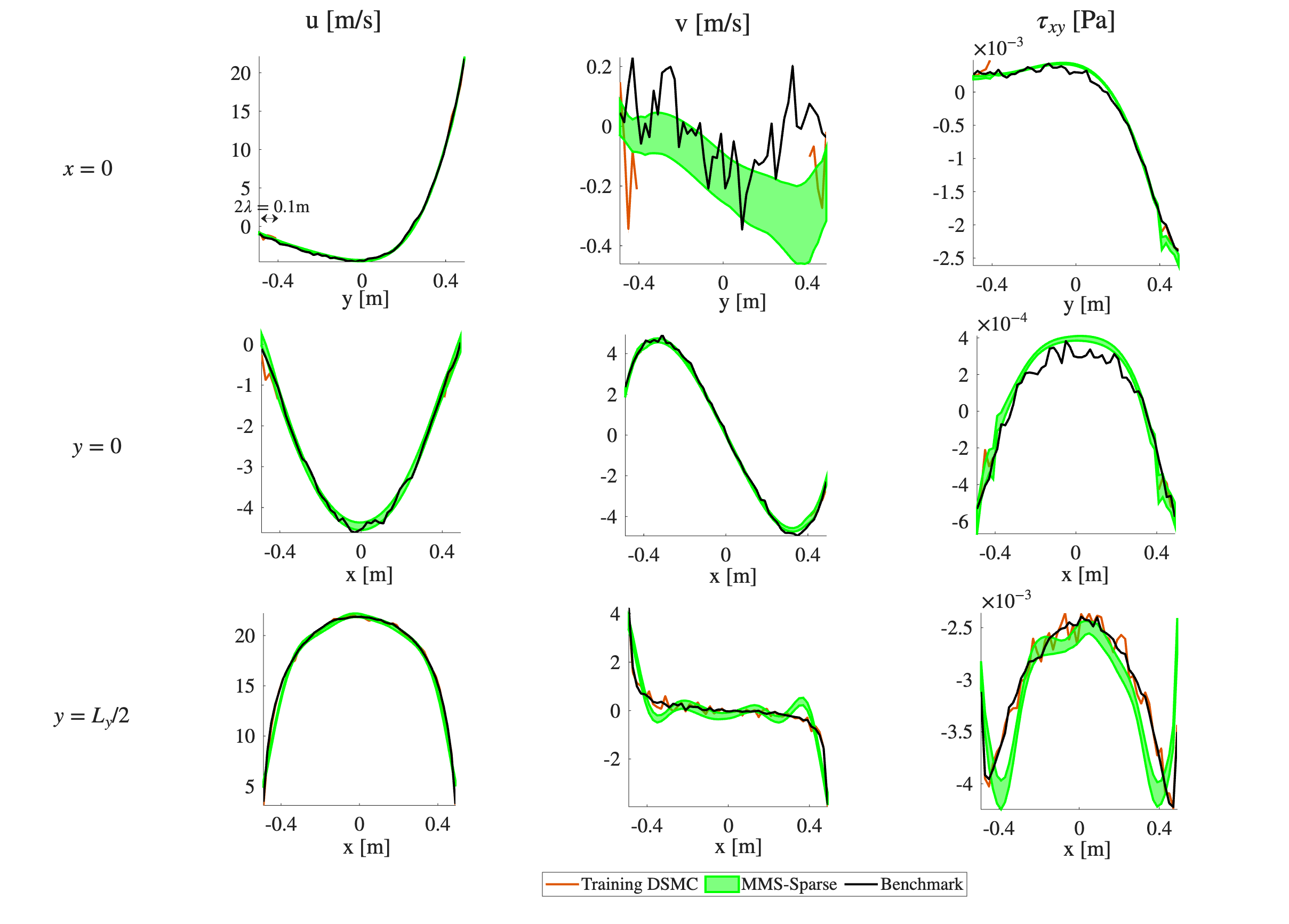}
	\centering 
	\caption{ Comparison at the centrelines and upper wall between training DSMC, MMS-Sparse with pseudo-hybrid approach, and benchmark DSMC at $Kn = 0.05$ $\&$ $M = 0.1$. }
	\label{fig:JointPlot_MMS_Hybrid_2Lambda_10Percent_Kn_0_05_M_0_1_2_SubDomain_CredibleIntervals}
\end{figure}

The $95 \%$ credible intervals for this case are shown in Figure \ref{fig:JointPlot_MMS_Hybrid_2Lambda_10Percent_Kn_0_05_M_0_1_2_SubDomain_CredibleIntervals}. For the flow velocity, we find that the results are good and similar to that observed in the proof-of-concept (see Figure \ref{fig:JointPlot_MMS_Hybrid_10Percent_Kn_0_05_M_0_1_2_WholeDomain_CredibleIntervals}). For the shear stress, however, the results are noticeably different from the proof-of-concept, with agreement compared to the benchmark less. For example, the shear stress, with the pseudo-hybrid approach, appears to be overestimated at the centreline midpoints, with the benchmark results outside the $95\%$ credible bounds. 

\subsection{Transitional regime}

\begin{figure}
	\centering
	\includegraphics[scale=0.35]{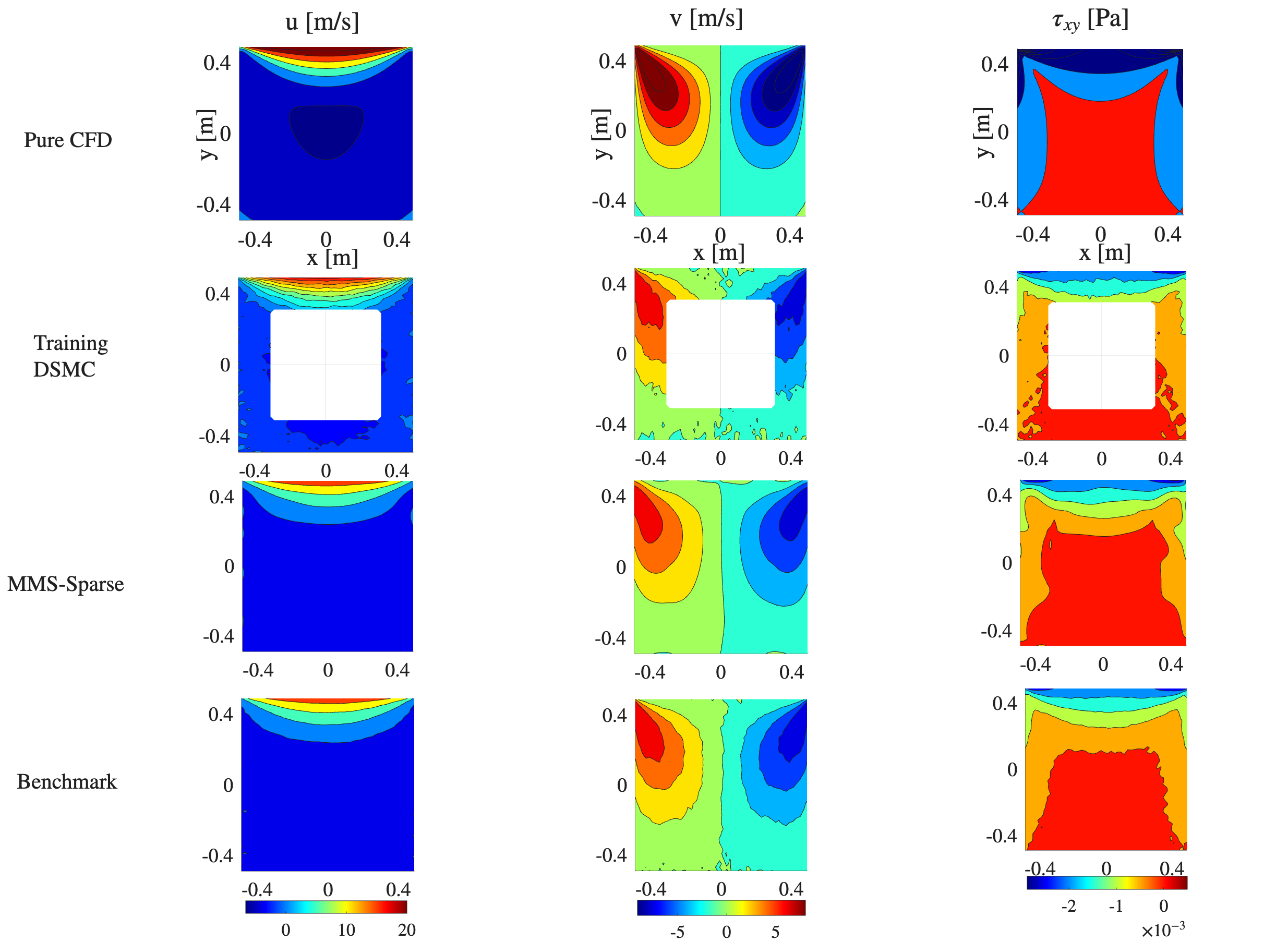}
	\caption{ Comparison between Pure CFD, training DSMC, MMS-Sparse with pseudo-hybrid approach, and benchmark DSMC at $Kn = 0.1$ $\&$ $M = 0.1$. }
	\label{fig:JointPlot_MMS_Hybrid_2Lambda_10Percent_Kn_0_1_M_0_1}
\end{figure}

Increasing the Knudsen number to $Kn = 0.1$, the comparison is shown in Figure \ref{fig:JointPlot_MMS_Hybrid_2Lambda_10Percent_Kn_0_1_M_0_1}. As with $Kn = 0.05$, we find good agreement for the flow velocity while there are, once again, deviations for the shear stress estimates, particularly towards the upper corners of the domain. Nevertheless, the pseudo-hybrid approach presents an improvement in estimating the macroscopic quantities compared to Pure CFD. \\

\begin{figure}
	\includegraphics[scale=0.35]{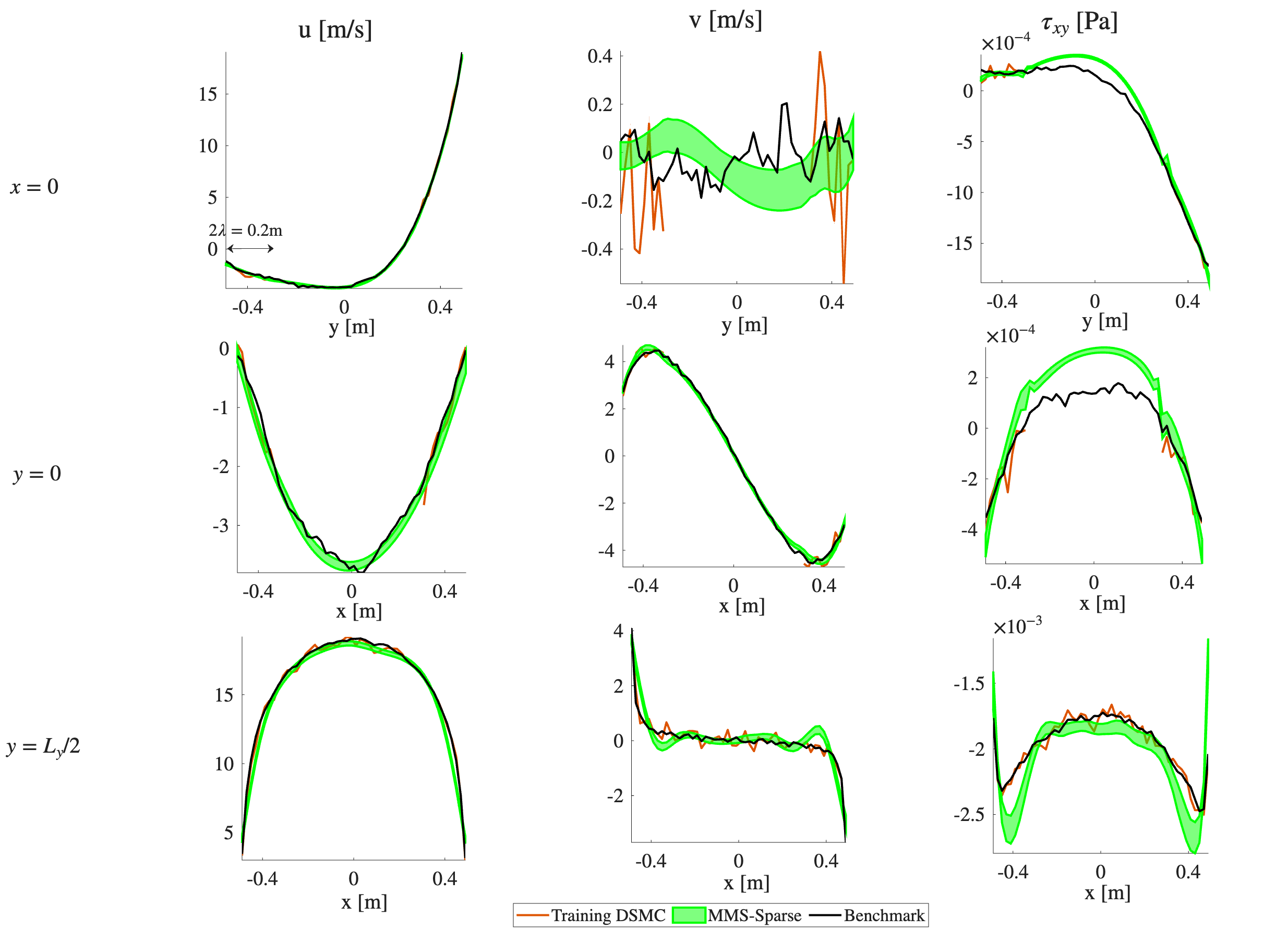}
	\centering 
	\caption{ Comparison at the centrelines and upper wall between training DSMC, MMS-Sparse with pseudo-hybrid approach, and benchmark DSMC at $Kn = 0.1$ $\&$ $M = 0.1$.  }
	\label{fig:JointPlot_MMS_Hybrid_2Lambda_10Percent_Kn_0_1_M_0_1_SubDomain_CredibleIntervals}
\end{figure}

Figure \ref{fig:JointPlot_MMS_Hybrid_2Lambda_10Percent_Kn_0_1_M_0_1_SubDomain_CredibleIntervals} shows the $95 \%$ credible intervals. For the flow velocity, the results are generally good. For the shear stress, as at $Kn = 0.05$, clear deviations are present at the vertical and horizontal midpoints, with MMS-Sparse with the pseudo-hybrid approach providing overestimates. Here, we find once more that the benchmark results are outside the $95\%$ credible bounds.

\section{Discussion}
\label{sec:discussion}

In this paper, we applied the existing MMS-Sparse method to the steady-state, spatially two-dimensional LDC flow problem, where multilevel RBFs were used as the basis set for greater geometric flexibility. Extending previous works \cite{tatsios2025dsmc} \cite{chinnappan2025bayesian}, the method was applied to a spatially two-dimensional problem, where corrections were computed in two-dimensions and an open-source finite-volume solver (i.e. OpenFOAM) was used obtain CFD estimates. We considered the LDC as the test problem as it was the logical next step after studying spatially one-dimensional microchannel flows and is significantly more complex. We used sparse Bayesian learning, specifically the SSBL algorithm, to ensure our CFD estimates were noise-robust and obtained through automation. We solved N-S equations with stress corrections and Dirichlet boundary corrections to ensure our CFD estimates are reliable satisfying conservation laws for mass and momentum. The novelty in our DSMC-CFD coupling was from the use of multilevel RBFs for the regression analysis, where RBFs up to level 4 ($17 \times 17$ RBFs) were used to learn the smooth corrections. \\

From application of MMS-Sparse as a proof-of-concept, we found that estimates for the flow velocity and shear stress for all the considered cases are generally good, with our DSMC-CFD coupling robust in reducing noise for all the cases and providing significant improvements compared to Pure CFD (i.e. CFD with zero corrections). From our estimates, we did consistently see slight oscillations at the side walls for the x-component of the flow velocity. This is potentially due to `Runge’s phenomenon' associated to the use of RBFs on a finite interval \cite{boyd2010six}. To address this limitation in future work, we can place additional, modified RBFs near the boundaries to suppress the oscillations. We also obtained larger deviations, compared to benchmark, at $Kn = 5$. This is to be expected, since using incompressible N-S as the macro-model is a significant approximation for the high $Kn$ case. Nevertheless, as well as providing accurate estimates for many of the test cases, we found that our full-domain DSMC-CFD coupling approach offered in-principle computational speed ups over full-domain benchmark DSMC by a factor of approximately $1.6$. It should be noted, however, that the real computational savings would come from a true hybrid where the DSMC is restricted to near-wall regions. We note that the potential savings from such a true hybrid differs from what is demonstrated in this paper, in that simulations for the training and benchmark are both run to steady-state and the DSMC simulations are performed for the full domain. In addition to the in-principle computational savings, we also computed the root-mean-square-error (RMSE) for MMS-Sparse and training DSMC. Although not shown, we found, overall, that the RMSE values for MMS-Sparse were not significantly smaller than that found for training DSMC due to the benchmark DSMC data themselves containing sampling noise. \\

In order to test the methodology closer to a true hybrid, we also applied MMS-Sparse in a pseudo-hybrid sense. Here, we captured the constitutive corrections, via multilevel RBFs, at a distance $2\lambda$ from the walls and was significantly more complex than using the whole domain (i.e. proof-of-concept). As in the proof-of-concept, for small Knudsen numbers $Kn = 0.05$ $\&$ $0.1$, the estimates for the flow velocity were very good, providing again improvements compared to Pure CFD. However, we saw two minor issues relating to the shear stress. The first issue was that the estimates had localised deviations compared to the benchmark close to the upper corner points. Since previous studies on MMS and MMS-Sparse \cite{tatsios2025dsmc} \cite{chinnappan2025bayesian} did not report such deviations, we anticipate that this deviation may be due to the singular nature of the LDC problem associated to the corner points of the moving upper wall. The second issue was that deviations for the shear stress were also found at the midpoints ($y = 0$). This is more pronounced at $Kn = 0.1$, whereas the same behaviour is not observed even at higher $Kn = 0.5$ when full-domain DSMC is used. This issue is likely a consequence of the imposed zero corrections beyond a distance $2\lambda$ (pre-defined) from the walls in the pseudo-hybrid setup. More general non-equilibrium detection criteria, such as the Kolmogorov-Smirnov approach \cite{titov2011determining}, may help alleviate this issue. \\

In addition, it should be noted that our approach for DSMC-CFD coupling is case specific. Therefore, there is no immediate transferability of the learned corrections to different conditions. For each new problem, fresh DSMC data and retraining are required. Furthermore, while our multilevel RBF approach is presented for structured cartesian grids, it is also applicable to unstructured meshes. This is because the design matrix required for learning computes the spatial distance between the structured RBFs and the computational mesh (structured or unstructured). Nevertheless, application to unstructured meshes and more complex geometries may require careful placement and distribution of RBF centres, particularly near irregular or curved boundaries, including adaptation of the centre placement to body-fitted off-wall layers. The selection of appropriate RBF widths may also require consideration for highly non-uniform meshes. Finally, we acknowledge that our applicability of the multilevel RBFs approach is confined to the two-dimensional LDC flow problem. And so, any claims on general applicability for more complex systems remain potential. Demonstration of the method for more complex domains will be explored in future work.

\subsection{Future work}

While we have demonstrated MMS-Sparse for low-speed rarefied flows in this paper, the demonstration for compressible gas flow is the next step. This should be tackled in relation to further complex test cases, such as rarefied Taylor-Couette flow \cite{stefanov1993monte} and three-dimensional LDC problem \cite{wang2018oscillatory}. Furthermore, in the DSMC-CFD coupling, transferring CFD estimates back to the DSMC in a two-way coupling method is of importance. For more complex geometries and flow configurations, where the local flow regime may vary substantially across the domain, bidirectional exchange of information between the CFD and DSMC regions may become important for maintaining thermodynamic and interface consistency between the continuum and kinetic approaches. Integrating the present MMS-Sparse framework into such fully coupled, two-way configurations could reduce the computational cost by reducing the DSMC sampling required to obtain sufficiently low-noise solutions and by replacing a larger portion of the DSMC domain with DSMC-informed stress corrections. Extension of the present framework to fully coupled configurations is left for future work. And so, prior studies using two-way-coupling approaches (e.g. \cite{borg2013multiscale}) should be revisited with surrogate modelling in mind.  \\

In addition, further developments in the surrogate modelling should be performed. While the use of SSBL algorithm is fast in denoising and producing smooth corrections, we found that regression estimates can be over-fitting when the noise of the DSMC is very high (see \ref{sec:convergence-mms-sparse}). One strategy to resolve the over-fitted estimates, within the sparse Bayesian learning framework, is to suppose that the posterior distributions for $\bm{\alpha}$ and $\beta^{-1}$ are to be summarised by their modes (see \ref{sec:gamma-hyperprior}). Our surrogate models also assume uncorrelated Gaussian noise, which was not a reasonable assumption since we obtained quite narrow credible intervals in Sections \ref{sec:proof-of-concept} and \ref{sec:pseudo-hybrid-approach}. We can, therefore, explore extensions to include correlation using Bayesian multivariate linear regression \cite{box2011bayesian} to overcome the narrow credible intervals. This would potentially be beneficial for the square cavity cases at $Kn = 0.05$ $\&$  $0.1$ and $M = 0.1$, where the benchmark solution was found not to be within the credible bounds over some regions.

\section*{Acknowledgements}

The authors would like to thank Dr. Giorgos Tatsios and Dr. Nikos Vasileiadis for useful discussions. The authors gratefully acknowledge support from the Engineering and Physical Sciences Research Council (EPSRC), UK [grant number EP/V01207X/1]. The authors also acknowledge the use of High Performance Computing (HPC) facilities at University of Warwick.

\section*{Data statement}
Data and scripts supporting the results of this paper can be found in the following GitHub repository: \url{https://github.com/arshadkamal10/LDC}.

\printbibliography

\appendix

\section{The Pressure-Implicit with Splitting of Operators algorithm}
\label{sec:piso-algorithm}

For the macro-model, we use the icoFoam solver to solve the incompressible, transient N-S equations \eqref{eq:NavierStokesMomentum} - \eqref{eq:NavierStokesContinuity}, with Dirichlet boundary conditions for the flow velocity ${\bf u}$ and scalar pressure $p$. This solver uses the Pressure-Implicit with Splitting of Operators (PISO) algorithm, which we briefly describe this here. The reader is also referred to \cite{ferziger2002computational} for more detail. \\

The PISO algorithm is a Semi-Implicit Method for Pressure-Linked Equations (SIMPLE)-type algorithm, consisting of predictor and corrector steps needed to comply with the mass continuity equation. Steps of the algorithm can be summarised as follows:

\begin{enumerate}
	\item Set initial conditions for flow velocity and scalar pressure.
	\item Start procedure to compute next time step values. \label{Step2}
	\item Execute the PISO loop until updates in the corrected solution are sufficiently small: 
	\begin{enumerate}
		\item Solve the momentum equation \eqref{eq:NavierStokesMomentum} for the flow velocity found by finite-volume discretisation ({\bf Predictor Step}). 
		\item Given flow velocity and fluxes at the faces of the cells, solve the pressure-correction equation, found by discretising the continuity equation \eqref{eq:NavierStokesContinuity}, for a new scalar pressure solution ({\bf Corrector Step 1}).
		\item Correct flow velocity and scalar pressure further based on new pressure solution and the second pressure-correction equation found via the momentum and continuity equations \eqref{eq:NavierStokesMomentum} - \eqref{eq:NavierStokesContinuity} ({\bf Corrector Step 2}).
	\end{enumerate}
	\item Store converged flow velocity and scalar pressure values, as well as associated fluxes, at current time-step. 
	\item Terminate here if final time is reached. Otherwise, go to Step \ref{Step2} for the next time-step.
\end{enumerate}

\section{Gamma hyperpriors in sparse Bayesian learning}
\label{sec:gamma-hyperprior}

When the noise of the DSMC is very high, we found that the SSBL algorithm can produce over-fitted estimates that are unsuitable for use in a CFD model. Here, we briefly detail a strategy to overcome the over-fitted estimates within the sparse Bayesian framework. \\

Based on the original paper on the relevance vector machine (RVM) \cite{tipping2001sparse}, we assign gamma distributions to the hyperparameters $\bm{\alpha}$ and $\beta$ resulting in the following algorithm:

\begin{enumerate}
	\item Set threshold $w_{thres}$ for the weights $w_{i}$ based on known parameters (e.g., target vector $\hat{{\bf d}}$ and design matrix ${\bm \Psi}$): $w_{thres} = \hspace{1mm} 10^{-4}\times \vert\vert {\hat {\bf d}} \vert\vert/ \vert\vert {\bm \Psi} \vert\vert$. 
	\item Set also thresholds for convergence in $\bm{\alpha}$ and $\beta$ (i.e., $10^{-4}$).
	\item Initialise $\bm{\alpha}$ and $\beta$ from gamma distributions of the form $\Gamma(2, 2 / A_{i})$ and $\Gamma(2, 2 / A)$, where $A_{i}$ and $A$ are some large parameters. \label{GammaRVMStep2}
	\item Calculate mean ${\bf m}$ and covariance ${\bm \Sigma}$ as before, using equations \eqref{eq:MeanSSBL} and \eqref{eq:CovarianceSSBL}. \label{GammaAlgorithmStep3}
	\item Update $\bm{\alpha}$ and $\beta$, taking into account the gamma distribution parameters $a_{1} = a_{2} = 2$ and $b_{1} = 2 / A_{i}, b_{2} = 2 / A$:
	\begin{eqnarray}
	\alpha_{i}^{new} = \frac{\gamma_{i} + 2a_{1}}{m_{i}^{2} + 2b_{1}}, \\
	(\beta^{new})^{-1} = \frac{\vert\vert \hat{{\bf d}} - {\bm \Psi} {\bf m} \vert\vert^{2} + 2b_{2}}{N_{x}N_{y} - \sum_{i} \gamma_{i} + 2a_{2}}, 
	\end{eqnarray}
	where we have defined $\gamma_{i} = 1 - \alpha_{i}\Sigma_{ii}$ as in \cite{mackay1992bayesian}. \label{GammaRVMStep4}
	\item Prune components of $\bm{\alpha}$ whose corresponding mean $\vert m_{i} \vert < w_{thres}$. \label{GammaAlgorithmStep6}
	\item Repeat Step \ref{GammaAlgorithmStep3} to Step \ref{GammaAlgorithmStep6} until convergence criteria is met for ${\bm{\alpha}}$ and $\beta$.
\end{enumerate}

In this algorithm, in Steps \ref{GammaRVMStep2} and \ref{GammaRVMStep4}, we consider $A = 5\times 10^{2}$ for the inverse-variance $\beta$ and $A_{i} = 5\times 10^{2}\times i^{2}$ for level $i$ corresponding to $\alpha_{i}$. With these choices, the resulting posterior distributions are assumed to depend on the mode of the dataset \cite{gelman1995bayesian}. 

\subsection{Low-noise vs. high-noise}
\label{ssec:gamma-hyperprior}

\begin{figure}
	\centering
    \includegraphics[scale=0.35]{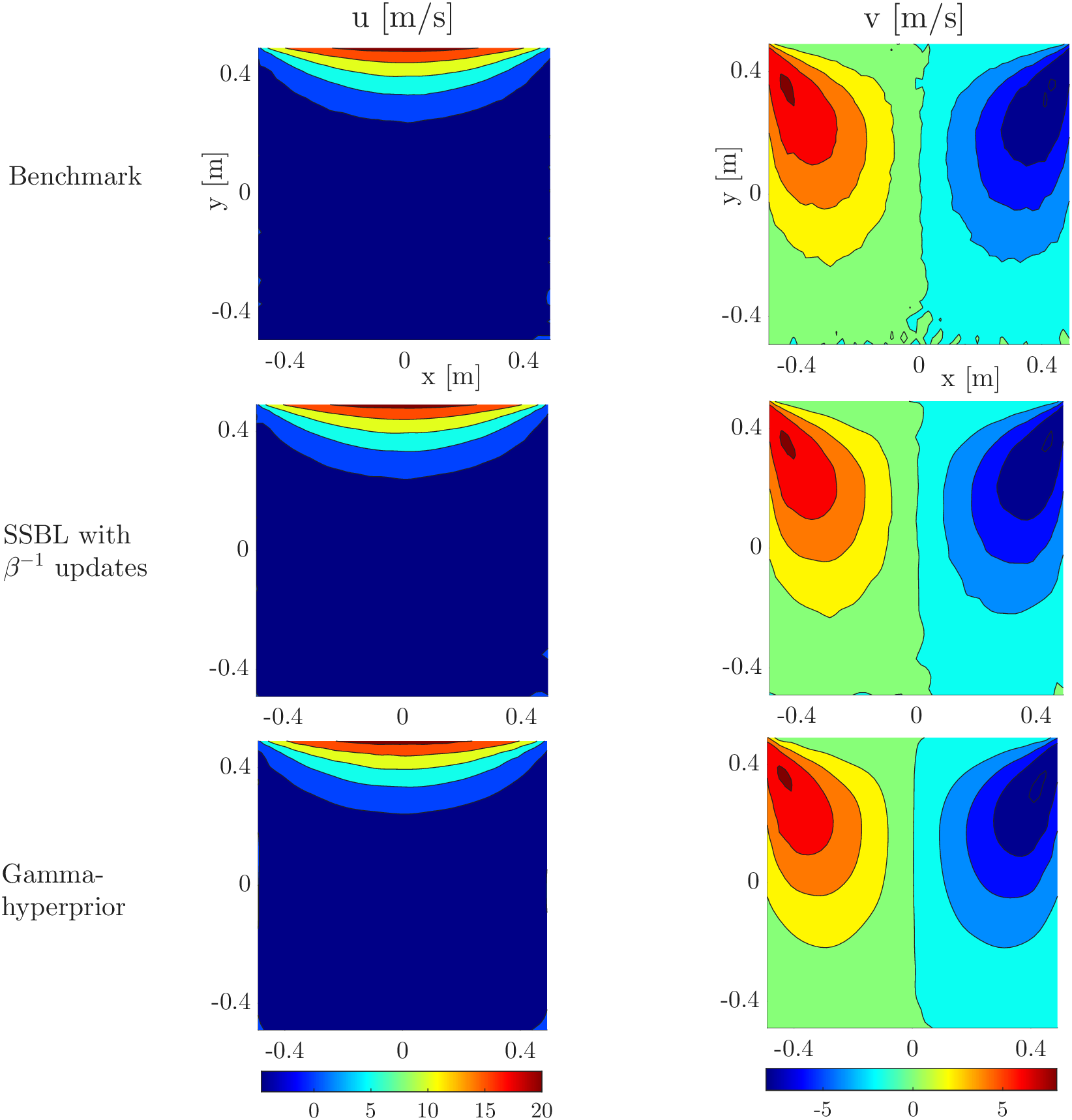}
	\caption{ Regression estimates using SSBL with $\beta^{-1}$ updates and gamma hyperpriors (both up to level 6) for low-noise benchmark DSMC data at $Kn = 0.05$ $\&$ $M = 0.1$. }
	\label{fig:Kn_0_05_M_0_1_Flat_vs_Gamma_Hyperprior_100_Percent_Benchmark}
\end{figure}

We consider both low-noise (benchmark DSMC), and high-noise at $Kn = 0.05$ and $M = 0.1$, and make a comparison between the SSBL algorithm and the gamma hyperprior algorithm. As a verification that the general RVM algorithm with a choice of gamma hyperpriors can reproduce close to noise-free data, we apply it to the benchmark DSMC data. As shown in Figure \ref{fig:Kn_0_05_M_0_1_Flat_vs_Gamma_Hyperprior_100_Percent_Benchmark}, the direct RVM reproduces the findings of the SSBL algorithm. \\

\begin{figure}
	\centering
    \includegraphics[scale=0.35]{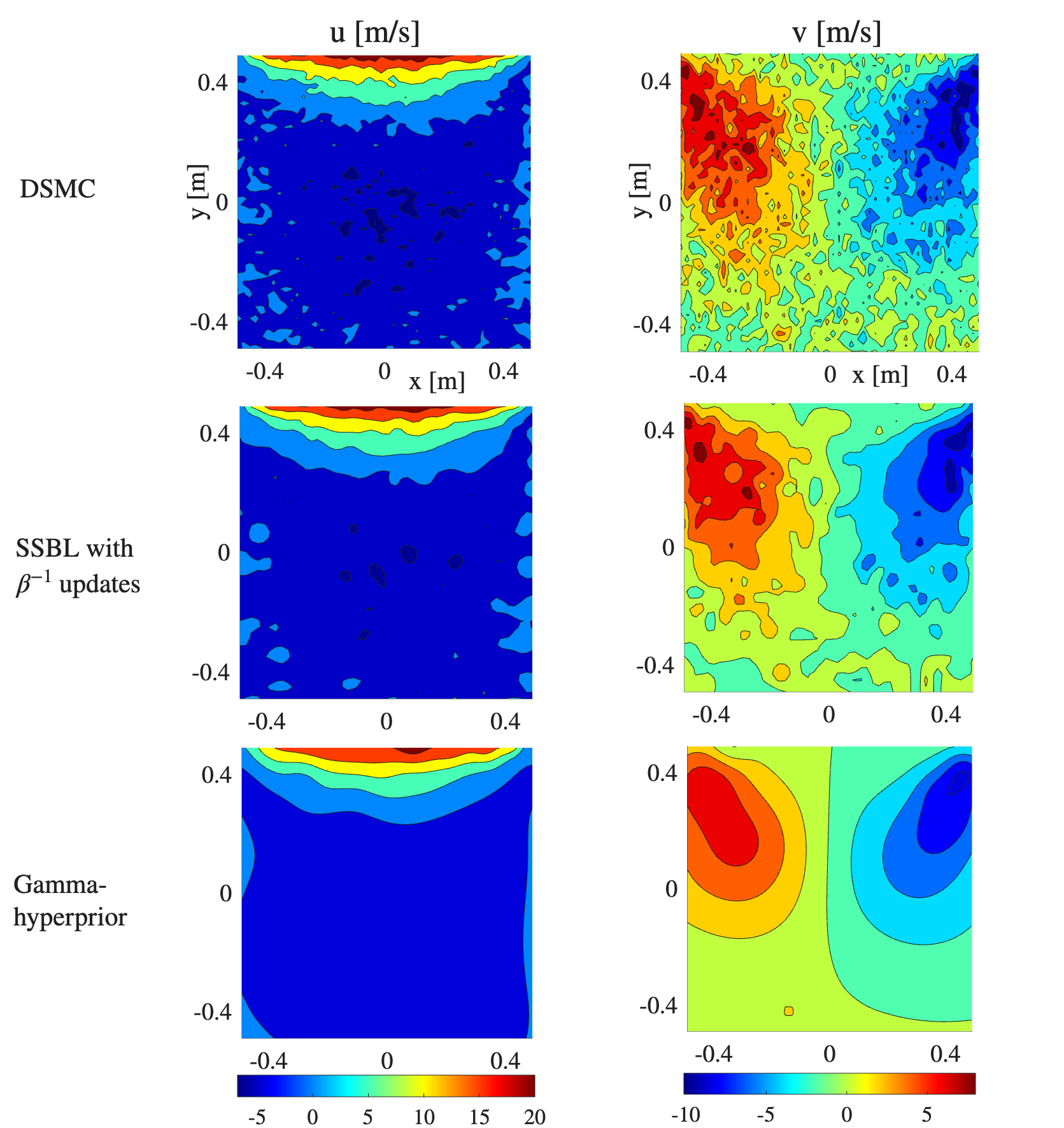}
	\caption{ Regression estimates using SSBL and gamma hyperpriors (both up to level 6) for high-noise DSMC data at $Kn = 0.05$ $\&$ $M = 0.1$. }
	\label{fig:Kn_0_05_M_0_1_Flat_vs_Gamma_Hyperprior_0_5Percent_Benchmark}
\end{figure}

We next examine high-noise data (two-hundredth of the aggregated benchmark data) and apply SSBL algorithm up to level 6. The comparison is shown in Figure \ref{fig:Kn_0_05_M_0_1_Flat_vs_Gamma_Hyperprior_0_5Percent_Benchmark}. For this high noise data, we see that higher-order features can be retained with the SSBL algorithm. By employing the more general RVM algorithm, we can, to some extent, filters out these higher-order features as one would like for use in a continuum model.  

\section{Fixing level for radial basis functions}
\label{sec:RBFs-level}

\begin{figure}
	\centering
    \includegraphics[scale=0.40]{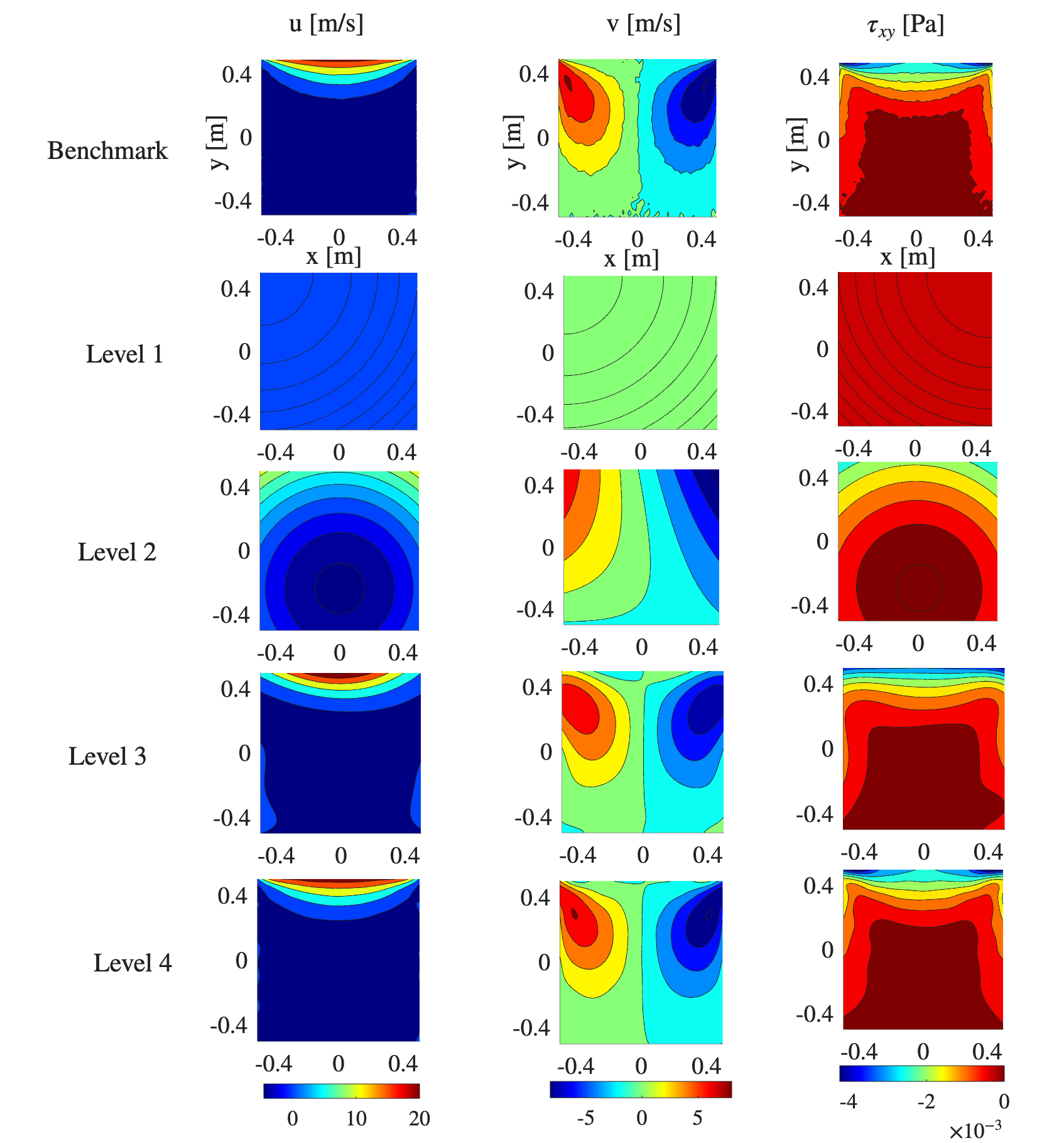}
	\caption{ Regression estimates using SSBL, with $\beta^{-1}$ updates, up to and including level 4 for Benchmark DSMC data at $Kn = 0.05$ $\&$ $M = 0.1$.  }
	\label{fig:Kn_0_05_M_0_1_Fixing_Level_100_Percent_Benchmark}
\end{figure}

To fix the number of RBFs, or precisely the level, we consider fitting using the SSBL algorithm to benchmark DSMC data at $Kn = 0.05$ $\&$ $M = 0.1$, with $u, v$ $\&$ $\tau_{xy}$ as the target variables. Figure \ref{fig:Kn_0_05_M_0_1_Fixing_Level_100_Percent_Benchmark} shows that considering RBFs up to level 4 (i.e., $17 \times 17 = 289$ RBFs) is sufficient to capture the required features of the target. Here, in order to satisfy the requirement that the reciprocal condition number of $\Psi$ is $> 10^{-12}$, we set $\sigma^{x}_{1:4} / (\Delta x)_{1:4} = \sigma^{y}_{1:4} / (\Delta y)_{1:4} \approx 1.8$.  

\section{Convergence of MMS-Sparse to Benchmark DSMC}
\label{sec:convergence-mms-sparse}

\begin{figure}
	\centering
    \includegraphics[scale=0.35]{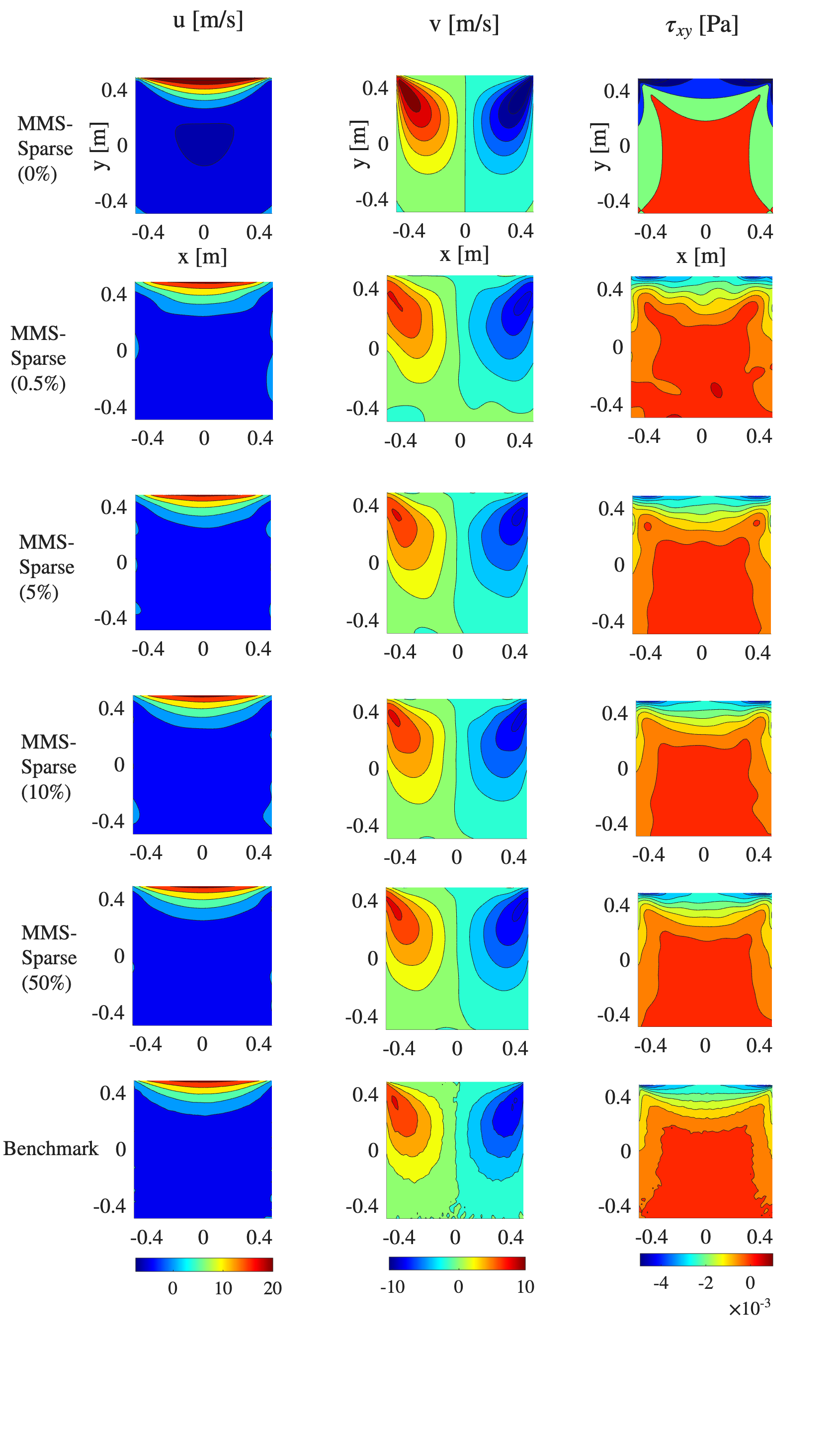}
	\caption{ MMS-Sparse with different percentages of the aggregated benchmark DSMC data at $Kn = 0.05$ $\&$ $M = 0.1$.  }
	\label{fig:AR-1-Kn-0p05-M-0p1-MMS-Sparse-Convergence-Columns}
\end{figure}

Figure \ref{fig:AR-1-Kn-0p05-M-0p1-MMS-Sparse-Convergence-Columns} shows results under MMS-Sparse for different percentages of the aggregated benchmark DSMC used as input. We see convergence in the profiles for the flow velocity and shear stress when increased up to $50\%$. We also note over-fitted estimates can be obtained for the shear stress when using DSMC data with very high noise (i.e. $0.5\%$ of the aggregated benchmark).

\end{document}